\documentclass[10pt]{IEEEtran}
\pdfoutput=1

\usepackage{graphicx}
\usepackage{hyperref}
\usepackage[utf8]{inputenc}
\usepackage{listings}
\usepackage[table]{xcolor}
\usepackage{pdfpages}

\usepackage{amsmath}
\usepackage{newtxtext}
\usepackage{newtxmath}
\usepackage{latexsym}
\usepackage{bm}
\usepackage{algorithm}
\usepackage{algorithmic}

\usepackage{xcolor}
\usepackage{soul}

\DeclareMathOperator*{\argmin}{arg\,min}

\hypersetup{colorlinks=true,citecolor=[rgb]{0,0.4,0}}

\title{Highly Versatile FPGA-Implemented Cyber Coherent Ising Machine}

\author{\IEEEauthorblockN{Toru Aonishi\IEEEauthorrefmark{1}\IEEEauthorrefmark{3}, Tatsuya Nagasawa\IEEEauthorrefmark{2}, Toshiyuki Koizumi\IEEEauthorrefmark{2}, Mastiyage Don Sudeera Hasaranga Gunathilaka\IEEEauthorrefmark{3}, Kazushi Mimura\IEEEauthorrefmark{4}, Masato Okada\IEEEauthorrefmark{1}, Satoshi Kako\IEEEauthorrefmark{5}, Yoshihisa Yamamoto\IEEEauthorrefmark{5}}

\IEEEauthorblockA{\IEEEauthorrefmark{1}\textit{Graduate School of Frontier Sciences, The University of Tokyo} 5-1-5 Kashiwanoha, Kashiwa-shi, Chiba 277-8561, JAPAN}

\IEEEauthorblockA{\IEEEauthorrefmark{2}\textit{D-CLUE Technologies Co.,Ltd.} 4F KAKiYA Building, 2-7-17 Shin-Yokohama, Kohoku-ku, Yokohama, Kanagawa 222-0033, JAPAN}

\IEEEauthorblockA{\IEEEauthorrefmark{3}\textit{School of Computing, Tokyo Institute of Technology} 4259-J2-38 Nagatsuda, Midori-ku, Yokohama, Kanagawa 226-8502, Japan}

\IEEEauthorblockA{\IEEEauthorrefmark{4}\textit{Graduate School of Information Sciences, Hiroshima City University} 3-4-1 Ohtsukahigashi, Asaminami-ku, Hiroshima 731-3194 Japan}

\IEEEauthorblockA{\IEEEauthorrefmark{5}\textit{Physics and Informatics Laboratories, NTT Research Inc.} 940 Stewart Dr, Sunnyvale, CA, 94085, USA}
}

\begin{document}
\maketitle

\begin{abstract}
In recent years, quantum Ising machines have drawn a lot of attention, but due to physical implementation constraints, it has been difficult to achieve dense coupling, such as full coupling with sufficient spins to handle practical large-scale applications. Consequently, classically computable equations have been derived from quantum master equations for these quantum Ising machines. Parallel implementations of these algorithms using field-programmable gate arrays (FPGAs) have been used to rapidly find solutions to these problems on a scale that is difficult to achieve in physical systems. 
We have developed an FPGA implemented cyber coherent Ising machine (cyber CIM) that is much more versatile than previous implementations using FPGAs. 
Our architecture is versatile since it can be applied to the open-loop CIM, which was proposed when CIM research began, to the closed-loop CIM, which has been used recently, as well as to Jacobi successive over-relaxation method.
By modifying the sequence control code for the calculation control module, other algorithms such as Simulated Bifurcation (SB) can also be implemented.
Earlier research on large-scale FPGA implementations of SB and CIM used binary or ternary discrete values for connections, whereas the cyber CIM used single-precision floating-point format (FP32) values.
Also, the cyber CIM utilized Zeeman terms that were represented as FP32, which were not present in other large-scale FPGA systems.
Our implementation with continuous interaction realizes $N=4096 $ on a single FPGA, comparable to the single-FPGA implementation of SB with binary interactions, with $N=4096$. The cyber CIM enables applications such as code division multiple access multi-user detector and L0-norm regularization-based compressed sensing which were not possible with earlier FPGA systems, while enabling superior calculation speeds, more than ten times faster than a GPU implementation. 
The calculation speed can be further improved by increasing parallelism, such as through clustering.

\end{abstract}

\section{Introduction}
Combinatorial optimization problems involve finding a solution for a combination of discrete parameters that maximizes (or minimizes) a particular evaluation index within given conditions.
The combination that meets all constraints does not exist for most real-world problems. Under such conditions, as the size of the problem increases, it becomes more difficult to find an optimal combination. According to computational complexity theory, such problems are NP-hard \cite{RN1161}, and require computing time that scales exponentially with the problem size. Combinatorial optimization problems include various real-world problems such as resource allocation \cite{RN1175}, scheduling \cite{RN1171}, portfolio optimization \cite{RN1173}, traffic flow optimization \cite{RN1174}, drug discovery \cite{RN1170,RN1169}, and machine learning \cite{RN1168,RN1167,RN1166,RN1165}. Because of the importance of these problems, there has been a strong demand for high-speed solution searches for combinatorial optimization problems, and thus a series of heuristics have been proposed to solve these problems \cite{RN1162,RN1164,RN1163}.

Ising models are composed of microscopic elements called Ising spins, which can take either of two states: up or down. This model can form complex macroscopic states such as glass states in which ferromagnetic and antiferromagnetic interactions (force the spins to align in the same and opposite directions respectively) coexist. Many combinatorial optimization problems, such as quadratic unconstrained binary optimization (QUBO), can be mapped to the task of finding the ground state of an Ising Hamiltonian. Consequently, to perform high-speed solution searches for the combinatorial optimization problems described above, Ising machines, which are dedicated hardware for rapidly searching for the ground state of Ising Hamiltonians, have been actively researched in recent years.
In particular, quantum Ising machines have garnered increasing interest due to their possession of quantum attributes that could offer potential solutions to the challenges inherent in large-scale combinatorial optimization problems.

Several quantum Ising machines have been proposed. This includes D-Wave systems\cite{RN1182,RN1181}, which utilize superconducting quantum interferometers \cite{RN1180,RN1179,RN1178} to execute quantum annealing for adiabatic quantum computation\cite{RN1177,RN1176}. There is also a quantum bifurcation machine that has been proposed to execute non-dissipative quantum adiabatic optimization \cite{RN1155,RN1154,RN1185}. Its implementation can be realized through either two-photon-driven Kerr parametric oscillators \cite{RN1183} or superconducting quantum circuits \cite{RN1184}. Moreover, Coherent Ising Machines (CIMs) represent a dissipative optical-parametric-oscillation (OPO) network leveraging the minimum gain principle for solving optimization problems \cite{RN1190,RN1188,RN825,RN924,RN1156,RN1187,RN1186,RN1140}.
In physically implemented systems based on superconducting or nonlinear quantum optical phenomena, the attainment of a large number of spins and dense interconnections presents considerable technical hurdles. This limitation primarily stems from the difficulty in physical wiring among a large number of spins. Consequently, there have been efforts to derive classically computable mathematical formulas from theoretical models of quantum Ising machines, often articulated in quantum master equations, Schrödinger equations, and related frameworks. Then use them to solve combinatorial optimization problems. 
  By deploying these algorithms in parallel on graphics processing units (GPUs) \cite{RN1050,RN1141,RN1142} and field-programmable gate arrays (FPGAs), it has become feasible to expeditiously solve problems necessitating extensive system sizes and dense interconnections. These are challenges that would be cumbersome to address in a physical system. This paper focuses on parallel computing architectures on FPGAs, which can also be implemented with application-specific integrated circuits (ASICs). Simulated Bifurcation (SB) has been proposed as a classically computable model for quantum bifurcation machines \cite{RN1050,RN1160}. SB has been implemented on a single FPGA \cite{RN1050,RN1139,RN1135} and on a large-scale multi-node FPGA cluster \cite{RN1135,RN1192}. $N=4096$ is implemented in a single FPGA \cite{RN1139,RN1135}, and $N=32,768$ is implemented in an 8-node FPGA cluster \cite{RN1192}. It is claimed that the FPGA-implemented SB is more efficient at solving Max-cut than the physical CIM system for $N=2000$ \cite{RN1050,RN1135}. In addition, attempts have been made to apply the FPGA-implemented SB to portfolio problems \cite{RN1128,RN1130,RN1131}. 
  
  In contrast, a stochastic differential equation (SDE) describing the amplitudes of optical-parametric-oscillation (OPO) pulses, derived from the quantum master equation using either the Wigner representation or the Positive-\textit{P} representation, serves as a computationally tractable model for Coherent Ising Machines (CIM).\cite{RN1195,RN1193,RN1194,RN1201,RN1202}. 
  In the CIM literature, a model integrating feedback kinetics to enable chaotic amplitude control (CAC) is called a closed-loop model \cite{RN1196,RN1153,RN1200,RN1158,RN1159,RN1143}, while a model lacking such kinetics is denoted as an open-loop model \cite{RN1196}. The closed-loop model has been realized on an FPGA \cite{RN1133}. Despite its calculation speed being inferior to that of FPGA-implemented SB, it has been shown to yield a higher success rate in solution finding. Furthermore, it has demonstrated a small slope in solution search time with respect to the system size on a logarithmic scale \cite{RN1133}.

The classically computable models have been implemented on FPGAs as described above. However, as the system size $N$ increases, the logic capacity and internal memory capacity becomes insufficient to implement such models in parallel on FPGA. Due to this capacity problem and problems with the models themselves as described below, for larger system sizes than $N\geq 1024$, their functions are restricted as follows \cite{RN1050,RN1139,RN1135,RN1192,RN1133}.
\begin{enumerate}
\item Connections are low-bit discrete values, and not continuous real values.
\item The Zeeman terms are not configurable.
\item Handling of either QUBO or Ising Hamiltonian optimization.
\item The implemented algorithm cannot be changed.
\end{enumerate}
(1) The most expensive process in the calculation of the models is the multiplication of the $N\times N$ coupling matrix with the $N$-dimensional spin-state vector. With both CIM and SB, when the system size is large, binary or ternary value connections are used to reduce the consumption of memory and logic resources and to improve their calculation speeds \cite{RN1050,RN1139,RN1135,RN1192,RN1133}. Continuous real-valued connections are needed to handle real-world problems. 
(2) The Zeeman terms play a crucial role in most optimization problems pertinent to signal processing applications, including compressed sensing \cite{RN1141,RN1142}, code division multiple access (CDMA) multi-user detectors \cite{RN1144}, multi-input multi-output (MIMO) systems \cite{RN1149,RN1152,RN1151} and other problems such as portfolio optimization \cite{RN1128,RN1130,RN1131}. Given that the spin variables in the aforementioned models adopt continuous values (without adhering strictly to $+1$ or $-1$), there arises a discrepancy between the magnitudes of the Zeeman term and the interaction term, leading to a degradation in their performance \cite{RN1144}. Therefore, it is necessary to devise ways to handle the Zeeman term. In CIM, where solutions to this problem have recently been proposed \cite{RN1153,RN1143,RN1141,RN1142,RN1204}, the Zeeman term is not yet implemented in an FPGA system. In SB \cite{RN1160}, the diagonal elements of the interactions i.e. self-interactions are used as Zeeman terms. Since the interaction is binary as described in (1), the Zeeman term can only be configured with binary values \cite{RN1128}. 
(3) With the models using continuous valued spin variables, to handle the QUBO problem with Zeeman terms, it is necessary to configure a suitable local field that is different from the Ising optimization\cite{RN1141,RN1142}. There is no single architecture capable of performing both of these.
(4) The models update frequently. Depending on the problem, the appropriate model can also differ. Currently, every time the model changes, the architecture needs to be redesigned. The flexibility of being able to change the model on FPGA is required. 

In this research, we have developed a highly versatile FPGA-implemented cyber-CIM to overcome the issues described above. Our developed architecture is highly versatile, allowing three different algorithms, open-loop CIM \cite{RN1194,RN1202,RN1141} and closed-loop CIM \cite{RN1159,RN1143} for combinatorial optimization problems and Jacobi Successive Over-Relaxation (SOR) for quadratic optimization problems to be executed on the same modules. By rewriting the sequence control code in the calculation control module, other algorithms can also be executed, and in principle SB could also be executed. The coupling matrix, Zeeman terms and OPO amplitudes are all represented using single-precision floating-point format (FP32), so local field and time evolution (TE) calculations are operated with FP32. The system sizes realized on a single FPGA are $N=1024$, $N=2048$ and $N=4096$, which are comparable to $N=4096$ of SB implemented on a single FPGA\cite{RN1139,RN1135}. Our system has $1/4$ degree of parallelism of the single-FPGA SB implementation, so it requires approximately four-times the number of cycles for one step including the local field and TE calculations \cite{RN1139,RN1135}. As benchmarks to evaluate the FPGA system we developed, we used a CDMA multi-user detector, which is an example of optimization of Ising Hamiltonian including Zeeman terms \cite{RN1144}, and L0-norm based regularization compressed sensing (L0RBCS), which is an example of optimization of QUBO Hamiltonian including Zeeman terms \cite{RN1141,RN1142,RN1204}. Our architecture realizes an alternative optimization for L0RBCS by alternating repeatedly between open-loop CIM or closed-loop CIM, and Jacobi SOR, \cite{RN1141,RN1142,RN1204}. These problems cannot be executed on the earlier FPGA systems described above. Here, we assess the performance of the FPGA system we've developed by comparing the computed results and execution times for these problems against those obtained using GPUs.

\section{Implemented Algorithms}

\begin{figure}[tb]
\includegraphics[width=80mm]{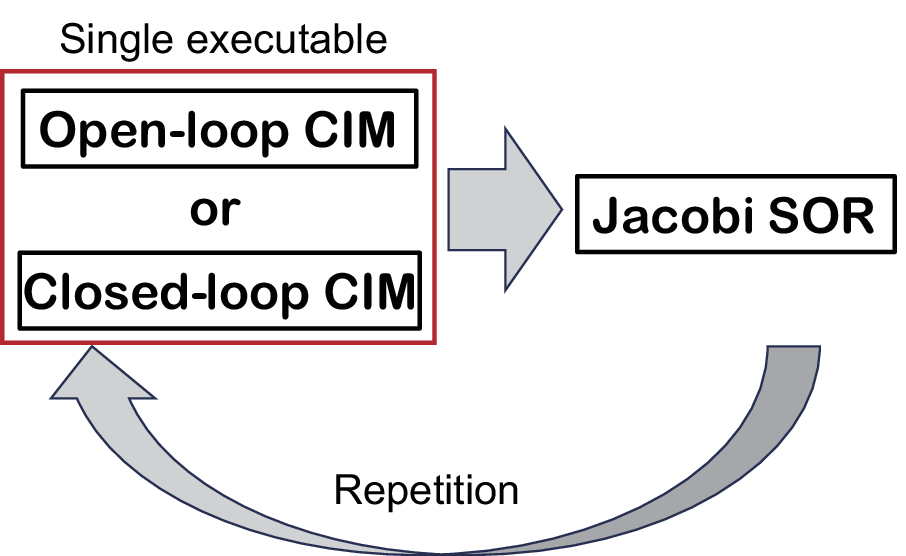}
\caption{Relationship among the implemented algorithms. L0-RBCS is realized by alternative execution between open-loop CIM or closed-loop CIM, and Jacobi SOR algorithms. It is also possible to execute only open-loop CIM or closed-loop CIM for various combinatorial optimization problems.}
\label{fig:algorithm_relation}
\end{figure}

We implemented the following three algorithms on the same FPGA architecture: open-loop CIM with Zeeman terms that does not perform amplitude control \cite{RN1194,RN1202,RN1141}, closed-loop CIM with Zeeman terms that performs chaotic amplitude control \cite{RN1159,RN1143,RN1204}, and Jacobi SOR that solves simultaneous equations. As shown in Fig. \ref{fig:algorithm_relation}, the architecture we have proposed can alternate repeatedly between open-loop CIM or closed-loop CIM, and Jacobi SOR, realizing an alternative optimization for L0RBCS \cite{RN1141,RN1142,RN1204}. It can also execute only open-loop CIM or closed-loop CIM, so it can be applied to various combinatorial optimization problems. We give an overview of each algorithm we implemented below.

\subsection{Supported Hamiltonians}

The algorithms we implemented support the following Ising Hamiltonian. 
\begin{eqnarray}
\mathscr{H} = -\frac{1}{2}\sum_{ij}^N J_{ij} \sigma_i \sigma_j - \sum_i^N g_i \sigma_i. \label{eq.hm1}
\end{eqnarray}
$\sigma_i$ is the $i$-th Ising spin taking either $+1$ or $-1$, $J_{ij}$ is the coupling strength between the $i$-th spin and $j$-th spin, and $g_i$ is the external field, which is called the Zeeman term.

The algorithms also support the following QUBO Hamiltonian. 
\begin{eqnarray}
\mathscr{H} = - \frac{1}{2}\sum_{ij}^N J_{ij} r_i r_j q_i q_j - \sum_i^N g_i r_i q_i + \lambda \sum_i^N q_i \label{eq.hm2}
\end{eqnarray}
$q_i$ is the $i$-th binary Potts spin taking either $0$ or $1$, $J_{ij}$ is the coupling strength between the $i$-th spin and the $j$-th spin, and $g_i$ is the Zeeman term. $\lambda$ is an external field corresponding to a threshold, common to each spin. $r_i$ is a real-value entry of the fixed auxiliary variable vector $r\in\Re^N$, which is configured according to the problem being solved. For most combinatorial optimization problems, it is required to set $r_i=1$ ($i=1,\cdots,N$). In contrast, if $q$ is fixed, the Hamiltonian in Eq. \ref{eq.hm2} with respect to $r$ becomes a quadratic function for $r$.

Both the open-loop CIM and closed-loop CIM described below perform the optimization of the above Ising Hamiltonian and QUBO Hamiltonian. The Jacobi SOR described below performs the quadratic optimization of the Hamiltonian in Eq. \ref{eq.hm2} with respect to $r$ under fixed $q$.

\subsection{Open-Loop CIM}

CIM is an Ising machine composed of a network of optical parametric oscillators (OPOs). Here time-multiplexed OPO pulses propagate along a cavity created by a long-distance optical fiber ring. The phase and amplitude of each OPO pulse are measured, and the mutual coupling among OPO pulses is computed using an FPGA. The injection field, representing the mutual coupling, is then fed back to each OPO pulse. This is an architecture of a measurement feedback-type CIM \cite{RN825,RN924}. 

By expanding the density operator of the whole OPO network with the Wigner function, applying Ito’s rule to the resulting Fokker-Planck equation, and ignoring measurement noise in the measurement feedback, we obtain the following equation \cite{RN1195,RN1193,RN1194,RN1202}.

\begin{eqnarray}
\frac{dc_i}{dt} &=& (-1+p-a_i)c_i + K I_{i} + g_s \sqrt{a_i + 0.5} W_{1i}, \nonumber \\
\frac{ds_i}{dt} &=& (-1-p-a_i)s_i + g_s\sqrt{a_i + 0.5} W_{2i}, \nonumber \\
& &a_i = c_i^2 + s_i^2, \ \ \ i=1,\cdots, N, \label{eq.1}
\end{eqnarray}
where $c_i$ and $s_i$ are the in-phase and quadrature-phase amplitudes of the $i$-th OPO pulse, which are normalized by the saturation parameter, $g_s$. $I_{i}$ is the injection field corresponding to the mutual coupling. This term only has an in-phase component, because the injection field is only injected into the in-phase component of the target OPO pulse in physical CIM. $K$ is the gain coefficient of the injection field. $p$ is the normalized pump rate. $p=1$ corresponds to the oscillation threshold of a solitary OPO without the mutual coupling. If $p$ is above the oscillation threshold ($p\lesssim 1$), each of the OPO pulses is either in the $0$-phase state or $\pi$-phase state. The $0$-phase of an OPO pulse is assigned to an Ising-spin up-state, while the $\pi$-phase is assigned to the down-state. $g_s$ is the saturation parameter which determines the nonlinear increase (abrupt jump) of the photon number at the OPO threshold. The last terms of the upper and lower equations express the vacuum fluctuations injected from external reservoirs and the pump fluctuations coupled to the OPO system via gain saturation. $W_{1i}(t)$ and $W_{2i}(t)$ are independent real Gaussian noise processes satisfying $\left<W_{ki}(t)\right> = 0$ and $\left< W_{ki}(t) W_{lj}(t’)\right> = \delta_{kl} \delta_{ij} \delta(t-t’)$. 

To be able to apply this to both the Ising Hamiltonian (Eq. \ref{eq.hm1}) and the QUBO Hamiltonian (Eq. \ref{eq.hm2}), we give the injection field $I_{i}$ as follows.
\begin{eqnarray}
  I_i &=& F_\chi(h_i) - \eta \\
  h_i &=& \left\{ \begin{array}{l}
   \sum_{j=1(\neq i)}^N J_{ij} c_j + g_i \; \; ({\rm Ising})\\
   \sum_{j=1(\neq i)}^N J_{ij} r_j H(c_j) + g_i \; \; ({\rm QUBO})
  \end{array}\right.,\label{eq.2}
\end{eqnarray}
where $h_i$ is called the local field in statistical mechanics, and $g_i$ is the Zeeman term introduced in Eqs. \ref{eq.hm1} and \ref{eq.hm2}. The injection field $I_i$ is calculated based on the local field $h_i$. The upper definition of $h_i$ where the pulse amplitude takes its original continuous value is for the Ising Hamiltonian (Eq. \ref{eq.hm1}), and the lower definition of $h_i$ where the pulse amplitude is binarized as 0 or 1 by the Heaviside step function $H$ is for the QUBO Hamiltonian (Eq. \ref{eq.hm2}) \cite{RN1141,RN1142}. $r_i$ is a real-value entry of the fixed auxiliary variable vector $r\in\Re^N$ defined in Eq. \ref{eq.hm2}. 
$\eta$ denotes a threshold value, which, in the context of L0RBCS, is associated with $\lambda$ in Eq. \ref{eq.hm2} via $\eta=\sqrt{2\lambda}$ \cite{RN1141}. In most other combinatorial optimization problems, it is required to set $\eta=0$ and $r_i=1$ ($i=1, \cdots, N$). $F_\chi$ is the function defined as follows. 
\begin{eqnarray}
  F_\chi(h) = \left\{ \begin{array}{l}
  h \;\;\; ( \chi = {\rm identity} ) \\
  |h| \;\; ( \chi = {\rm absolute} )
  \end{array}\right., \nonumber
\end{eqnarray}
For L0RBCS, the absolute value function ($\chi = {\rm absolute}$) is required. The reason that the absolute value function is required is discussed in our previous paper \cite{RN1141}. In most other combinatorial optimization problems, it is required to use the identity function ($\chi = {\rm identity}$).

Here, we solve the open-loop CIM stochastic differential equation using the Euler-Maruyama method. This numerical algorithm is shown in Algorithm \ref{alg1}.

\begin{algorithm}
\caption{Open-loop CIM}
\label{alg1}
\begin{algorithmic}[1]
\REQUIRE J matrix: $J\in\Re^{N\times N}$, Zeeman term: $g\in\Re^{N}$, Pump rate sequence: $p\in\Re^{N_{step}}$, Threshold: $\eta\in\Re$, Saturation parameter: $g_s\in\Re$, Gain: $K$, Index for $F_\chi\left(h\right)=|h|$ or $h$: $\chi$, Auxiliary variable for QUBO: $r\in\Re^N$.
\STATE $c\leftarrow 0$, $s\leftarrow 0$.
\STATE $\mu\leftarrow r$, $\sigma\leftarrow 0$ (if QUBO) or $\mu\leftarrow 0$, $\sigma\leftarrow 1$ (if Ising).
\FOR{$l=1$ to $N_{step}$}
\STATE // Matrix-Vector Multiplication (MM)

\FOR{$i=1$ to $N$}
\STATE $h_i \leftarrow \sum_{j=1(\neq i)}^{N} J_{ij} \sigma_j \mu_j + {g}_i$.
\ENDFOR
\STATE // Time Evolution (TE)
\FOR{$i=1$ to $N$}
\STATE $W_{i1} \leftarrow N(0,1), \ \ W_{i2} \leftarrow N(0,1)$
\STATE $a_i \leftarrow c_i^2+s_i^2, \ b_i\leftarrow \sqrt{1/2+a_i}$
\STATE $c_i \leftarrow c_i + \Delta t\left(\left(-1+p_l+a_i\right)c_i + K \left(F_\chi\left(h_i\right)-\eta\right)\right) +\sqrt{\Delta t}g_s b_i W_{i1}$
\STATE $s_i \leftarrow s_i + \Delta t \left(-1-p_l+a_i\right)s_i+\sqrt{\Delta t}g_s b_i W_{i2}$
\STATE $\mu_i \leftarrow c_i$ (if Ising) or $\sigma_i \leftarrow H(c_i)$ (if QUBO)
\ENDFOR
\ENDFOR
\end{algorithmic}
\end{algorithm}

\subsection{Closed-Loop CIM}

By integrating feedback kinetics for amplitude homogenous control with a SDE, derived from the quantum master equation for the entire OPO network using either the Wigner or Positive-P approximation, the closed-loop CIM model is derived.
Especially, the effectiveness of CAC feedback, which produces chaotic trajectories of OPO amplitudes, is attracting attention \cite{RN1196,RN1153,RN1200,RN1158,RN1159,RN1143,RN1204}. In the zero quantum noise limit, a simple deterministic differential equation called the mean-field model is obtained \cite{RN1143}. Due to the parametric oscillation, in-phase amplitude components are amplified, while quadrature-phase amplitude components are anti-amplified, so the in-phase amplitude components becomes dominant in the SDEs for CIMs. Therefore, the mean field model of the close-loop CIM is described by the equation for $c_i$ in Eq. \ref{eq.1} in the limit of $g_s \rightarrow 0$ and the CAC feedback kinetic equation \cite{RN1158,RN1159,RN1143,RN1204}: 
\begin{eqnarray}
\frac{dc_i}{dt} &=& (-1+p-a_i)c_i + K e_i I_{i}, \nonumber \\
\frac{de_i}{dt} &=& \beta(\tau - a_i) e_i, \nonumber \\
& &a_i = c_i^2, \ \ \ i=1,\cdots, N, \label{eq.4}
\end{eqnarray}
where $c_i$ is the in-phase amplitude of the $i$-th OPO pulse, $I_{i}$ is the injection field, $K$ is the gain coefficient of the injection field, and $p$ is the pump rate. $e_i$ is the feedback error to keep the pulse amplitudes homogeneous. $\tau$ is the target value for the squared amplitude $a_i$, and $\beta$ is the rate of exponential growth or decline for $e_i$.  It has been reported that forcefully trying to equalize the amplitudes of the system to a target amplitude may result in a chaotic behavior in the system which may result in escaping from local minima in the energy landscape \cite{RN1196,RN1153,RN1200,RN1158,RN1159,RN1143}. 

To enable this model to be applied to both Ising Hamiltonians (Eq. \ref{eq.hm1}) and QUBO Hamiltonians, (Eq. \ref{eq.hm2}), the injection field $I_{i}$ is given as follows.
\begin{eqnarray}
  I_i &=& r_i h_i - \lambda \\
  h_i &=& \left\{ \begin{array}{l}
   \sum_{j=1(\neq i)}^N J_{ij} c_j + g_i \; \; ({\rm Ising})\\
   \sum_{j=1(\neq i)}^N J_{ij} r_j H(c_j) + g_i \; \; ({\rm QUBO})
  \end{array}\right..\label{eq.5}
\end{eqnarray}
Here, $h_i$ is called the local field in statistical mechanics, and $g_i$ is the Zeeman term introduced in Eqs. \ref{eq.hm1} and \ref{eq.hm2}. $r_i$ is a real-value entry of the fixed auxiliary variable vector $r\in\Re^N$ defined in Eq. \ref{eq.hm2}. For closed-loop CIM, $F_\chi$ is not introduced, and the injection field $I_i$ is defined as the product of $h_i$ with $r_i$ \cite{RN1142,RN1204}. The upper definition of $h_i$ where the pulse amplitude takes its original continuous value is for the Ising Hamiltonian (Eq. \ref{eq.hm1}), and the lower definition of $h_i$ where the pulse amplitude is binarized as 0 or 1 by the Heaviside step function $H$ is for the QUBO Hamiltonian (Eq. \ref{eq.hm2}) \cite{RN1204}. $\lambda$ is a threshold, corresponding to $\lambda$ in Eq. \ref{eq.hm2}. For L0RBCS, $\lambda$ is related to $\eta$ in Eq. \ref{eq.2} by $\eta=\sqrt{2\lambda}$ \cite{RN1141}. In most other combinatorial optimization problems, it is required to set $\lambda=0$ and $r_i=1$ ($i=1, \cdots, N$). 

Here, we solve the differential equation for the closed-loop CIM using the Euler method. The numerical algorithm used to solve the differential equation is shown in Algorithm \ref{alg3}. Initial values shown in Algorithm \ref{alg3} were set according to our previous research \cite{RN1143,RN1204}.

\begin{algorithm}
\caption{Closed-loop CIM}
\label{alg3}
\begin{algorithmic}[1]
\REQUIRE J matrix: $J\in\Re^{N\times N}$, Zeeman terms: $g\in\Re^{N}$, Pump rate sequence: $p\in\Re^{N_{step}}$, Gain: $K$, Threshold: $\lambda\in\Re$, Target amplitude: $\tau\in\Re$, Feedback coefficient: $\beta\in\Re$, Auxiliary variable for QUBO: $r\in\Re^N$.
\STATE $c\leftarrow i.i.d. N(0,0.02)$, $e\leftarrow 1$.
\STATE $\mu\leftarrow r$, $\sigma\leftarrow 0$ (if QUBO) or $r\leftarrow 1$, $\mu\leftarrow 0$, $\sigma\leftarrow 1$ (if Ising).
\FOR{$l=1$ to $N_{step}$}
\STATE // Matrix-Vector Multiplication (MM)
\FOR{$i=1$ to $N$}
\STATE $h_i \leftarrow \sum_{j=1(\neq i)}^{N} J_{ij} \sigma_j \mu_j + {g}_i$.
\ENDFOR
\STATE // Time Evolution (TE)
\FOR{$i=1$ to $N$}

\STATE $a_i \leftarrow c_i^2$.
\STATE $c_i \leftarrow c_i + \Delta t\left(\left(-1+p_l+a_i\right)c_i + K e_i \left(r_i h_i - \lambda\right)\right)$.
\STATE $e_i \leftarrow e_i + \Delta t \beta \left(\tau - a_i\right)e_i$.
\STATE $\mu_i \leftarrow c_i$ (if Ising) or $\sigma_i \leftarrow H(c_i)$ (if QUBO).
\ENDFOR
\ENDFOR
\end{algorithmic}
\end{algorithm}

\subsection{Jacobi SOR}

The Hamiltonian in Eq. \ref{eq.hm2} with respect to $r$ if $q$ is fixed becomes a quadratic function for $r$. The optimal solution of the Hamiltonian with respect to $r$ is the same as a solution of the following simultaneous equations.
\begin{eqnarray}
\begin{bmatrix}
   J_{11} & J_{12} & \cdots & J_{1N}\\
   J_{21} & J_{22} & \cdots & J_{2N}\\
   \vdots & \vdots & \ddots & \vdots \\
   J_{N1} & J_{N2} & \cdots & J_{NN} 
\end{bmatrix}
 \begin{bmatrix}
  q_1 r_1 \\
  q_2 r_2\\
  \vdots \\
  q_N r_N
\end{bmatrix}
+
  \begin{bmatrix}
  g_1 \\
  g_2 \\
  \vdots \\
  g_N
\end{bmatrix}
=0
\end{eqnarray}
We seek to obtain $r=[r_1 \cdots r_N]^T$ that satisfies the simultaneous equations. $q_i$ is the $i$-the entry in $q\in \{0,1\}^N$ which are given externally. More detail is given in the description of L0RBCS below.

Jacobi Successive Over-Relaxation (Jacobi SOR), which is an extension of the Jacobi method by introducing a relaxation factor, is used to find the solution. The recursive update equation for solving the above simultaneous equation is as follows.
\begin{eqnarray}
r_i \leftarrow (1-\Delta t) r_i + \frac{\Delta t}{J_{ii}} \left(g_i + \sum_{j(\neq i)}J_{ij} q_j r_j\right)
\end{eqnarray}
Adjusting $\Delta t$ improves the numerical stability and the speed of convergence. The numerical algorithm of the Jacobi SOR is shown in Algorithm \ref{alg5}. In the algorithm, it is required to set $d_i=1/J_{ii}$.

\begin{algorithm}
\caption{Jacobi Successive Over-Relaxation (SOR)}
\label{alg5}
\begin{algorithmic}[1]
\REQUIRE J matrix: $J\in\Re^{N\times N}$, Zeeman terms: $g\in\Re^{N}$, Inverse of $J$ diagonal elements: $d\in\Re^N$, Auxiliary variable: $q\in\{0,1\}^N$.
\STATE $\sigma \leftarrow q$, $r \leftarrow 0$, $\mu \leftarrow 0$.
\FOR{$l=1$ to $N_{step}$}
\STATE // Matrix-Vector Multiplication (MM)
\FOR{$i=1$ to $N$}
\STATE $h_i \leftarrow \sum_{j=1(\neq i)}^{N} J_{ij} \sigma_j \mu_j + {g}_i$.
\ENDFOR
\STATE // Time Evolution (TE)
\FOR{$i=1$ to $N$}
\STATE $r_i \leftarrow r_i + \Delta t\left(-r_i + d_i h_i\right)$.
\STATE $\mu_i \leftarrow r_i$.
\ENDFOR
\ENDFOR
\end{algorithmic}
\end{algorithm}

\subsection{CDMA multiuser detector}

We briefly explain the CDMA multiuser detector applied to FPGA implemented cyber CIM. To formulate the CDMA multiuser detector, we introduce a model of multiple access, wherein $N$ users transmit information bits concurrently over a single communication channel \cite{RN1206,RN1205,RN1144}.
\begin{eqnarray}
  \begin{bmatrix}
  y^1 \\
  \vdots \\
  y^M
\end{bmatrix}
=\frac{1}{\sqrt{M}}
  \begin{bmatrix}
   \xi_{1}^1 & \xi_{2}^1 & \cdots & \xi_{N}^1\\
   \vdots & \vdots & \ddots & \vdots \\
   \xi_{1}^M & \xi_{2}^M & \cdots & \xi_{N}^M
\end{bmatrix}
 \begin{bmatrix}
  \sigma_1 \\
  \sigma_2\\
  \vdots \\
  \sigma_N
\end{bmatrix}
+
\begin{bmatrix}
  n^1 \\
  \vdots \\
  n^M
\end{bmatrix}.\label{eq.cdma1}
\end{eqnarray}
Here, $y=[y^1,\cdots,y^M]^T\in\Re^{M}$ is the $M$-bit received sequence. $\sigma=[\sigma_1,\cdots,\sigma_N]\in\{+1,-1\}^{N}$ is the information bits sent from $N$ users. $\xi_i=[\xi_i^1,\cdots,\xi_i^M]\in\{+1,-1\}^M$ is the $i$-th user’s spreading code with $M$ chips.  $\Xi=1/\sqrt{M}[(\xi_1)^T,\cdots,(\xi_N)^T]^T\in\Re^{M\times N}$ is the $M\times N$ spreading code matrix. $n=[n^1,\cdots,n^M]^T\in\Re^M$ is the additive Gaussian noise satisfying $\left<n^\mu\right>$=0 and $\left<n^\mu n^\nu\right>=\zeta^2 \delta_{\mu\nu}$.

This model represents the following process. The information bits $\sigma_1$ to $\sigma_N$ for each of $N$ users are sought to be transmitted over a single communication channel. Before transmitting, the spreading code $\xi_i$ with $M$ chips is assigned to each user, and multiplied with the information bit $\sigma_i$ for each user to perform the spreading encoding. The encoded sequences from all users are summed up, resulting in an $M$-bit sequence transmitted through a single Gaussian communication channel. The received $M$-bit sequence denoted $y$, is then obtained. Here, as an indicator to measure the difficulty of the problem, the spreading rate $\alpha$, defined as $\alpha=M/N$, is introduced. 

The objective of this problem is to simultaneously retrieve all user bits $\sigma$ from the received sequence $y$ under the condition that the spreading code matrix $\Xi$ is known. $\sigma$ is retrieved by solving the following optimization problem. 
\begin{eqnarray}
\hat{\sigma} = \argmin_{\sigma\in\{+1,-1\}^N} \left(\frac{1}{2}\left\|y - \Xi \sigma \right\|^2_2 \right). \label{eq.CDMA_optimization}
\end{eqnarray}
Here, $\hat{\sigma}$ is the inferred value of the information bits for all users. From Eq. \ref{eq.CDMA_optimization}, we can determine the Hamiltonian for the CDMA multi-user detector:
\begin{eqnarray}
 \mathcal{H} &=& -\frac{1}{2} \sum_{i,j=1}^N J_{ij} \sigma_i \sigma_j - \sum_{i=1}^N g_i \sigma_i, \label{eq.CDMA-Hamiltonian}\\
  J_{ij} &=& -\frac{1}{M}\sum_{\mu=1}^M\xi_i^\mu\xi_j^\mu, \ \ g_i = \frac{1}{\sqrt{M}}\sum_{\mu=1}^M \xi_i^\mu y^\mu. \label{eq. anti-hebb-cdma}
\end{eqnarray}
This Hamiltonian is the same as the Ising Hamiltonian in Eq. \ref{eq.hm1}.

\subsection{L0RBCS}

We briefly describe L0RBCS applied to FPGA implemented cyber CIM. 
To formulate L0RBCS, we introduce the following observation process model \cite{RN1141}.
\begin{eqnarray}
  \begin{bmatrix}
  y^1 \\
  \vdots \\
  y^M
\end{bmatrix}
=
  \begin{bmatrix}
   A_{1}^1 & A_{2}^1 & \cdots & A_{N}^1\\
   \vdots & \vdots & \ddots & \vdots \\
   A_{1}^M & A_{2}^M & \cdots & A_{N}^M
\end{bmatrix}
 \begin{bmatrix}
  x_1 \\
  x_2\\
  \vdots \\
  x_N
\end{bmatrix}
+
\begin{bmatrix}
  n^1 \\
  \vdots \\
  n^M
\end{bmatrix}.\label{eq.cs1}
\end{eqnarray}
Here, $y=[y^1,\cdots,y^M]^T\in\Re^M$ is the $M$-dimensional observed signal, $x=[x_1,\cdots,x_N]\in\Re^N$ is the $N$-dimensional true source signal, and $A=[A_i^\mu]\in\Re^{M\times N}$ is the $M \times N$ observation matrix. $n=[n^1,\cdots,n^M]^T\in\Re^M$ is the $M$-dimensional Gaussian observation noise satisfying $\left<n^\mu\right>=0$ and $\left<n^\mu n^\nu\right>=\zeta^2\delta_{\mu\nu}$.

Because the dimension $M$ of the observed signal $y$ is less than the dimension $N$ of the original signal $x$, as a prerequisite for compressed sensing, the simultaneous equations in Eq. \ref{eq.cs1} are indeterminate even if the observation noise $n$ is zero. The objective of compressed sensing is to recover the higher-dimensional source signal $x$ from the lower-dimensional observed signal $y$. Under the sparse condition for the source signal $x$ that the number of non-zero elements in $x$ is less than the dimension $M$ of the observed signal $y$, it is possible to recover $x$ from $y$ if the position of the non-zero elements in $x$ can be identified. Here, we introduce the compression rate $\alpha=M/N$, which is the ratio between the dimension $M$ of $y$ and the dimension $N$ of $x$, and also the sparseness, $a$, which is the ratio of the number of non-zero elements in $x$ and the dimension $N$ of $x$. 

We reconstruct the sparse source signal $x$ by solving the following optimization problem, which incorporates L0-norm regularization.
\begin{eqnarray}
x = \argmin_{x\in\mathbb{R}^N} \left(\frac{1}{2}\left\|y - A x \right\|^2_2 + \sum_k\frac{\gamma_k}{2}\left\|\Gamma_k x \right\|^2_2 + \lambda \left\|x \right\|_0 \right). \label{eq.L0-regularization-based-CS}
\end{eqnarray}
Here, $\sum_k\frac{\gamma_k}{2}\left\|\Gamma_k x \right\|^2_2$ is the L2-norm regularization term to improve the estimation accuracy \cite{RN1209}. $\Gamma_k$ is a linear operator matrix, e.g., discrete-differential matrix to detect changing points. The L0RBCS formulation in Eq. \ref{eq.L0-regularization-based-CS} can be reformulated into a two-fold optimization problem \cite{RN1207,RN1208}:
\begin{eqnarray}
  (\hat{r},\hat{q}) &=& \argmin_{q\in\{0,1\}^N} \argmin_{r\in\mathbb{R}^N} \left(\frac{1}{2}\left\|y - A \left(q \circ r\right) \right\|^2_2 \right. \nonumber\\
  & &\left.+ \sum_k\frac{\gamma_k}{2}\left\|\Gamma_k \left(q \circ r\right) \right\|^2_2 + \lambda \left\|q \right\|_0 \right). \label{eq.L0-regularization-based-CS-support}
\end{eqnarray}
Here, $\hat{r}$ is the estimated value of the $N$-dimensional source signal and each element $r_i$ in $r$ represents the real-number value of the $i$-th element in the source signal. $\hat{q}$ is the estimated value of a support vector, which represents the places of the non-zero elements in the $N$-dimensional source signal. The entry $q_i$ in $q$ takes either 0 or 1 to indicate whether the $i$-th element in the source signal is zero or non-zero. The symbol $\circ$ denotes the Hadamard (element-wise) product. From the elementwise representation of Eq. \ref{eq.L0-regularization-based-CS-support}, the Hamiltonian of L0RBCS can be written as
\begin{eqnarray}
  \mathcal{H} &=& -\frac{1}{2} \sum_{i,j=1}^N J_{ij} r_i r_j q_i q_j - \sum_{i=1}^N g_i r_i q_i + \lambda \sum_{i=1}^N q_i, \label{eq.L0-regularization-based-CS-Hamiltonian}\\
  J &=& -A^T A +\sum_k\gamma_k \Gamma_k^T\Gamma_k, \ \ g = A^T y. \label{eq.CS_zeeman}
\end{eqnarray}
This Hamiltonian is the same as the QUBO Hamiltonian in Eq. \ref{eq.hm2}. Therefore as described in the above sections, Algorithm \ref{alg1} (open-loop CIM) and Algorithm \ref{alg3} (closed-loop CIM) for QUBO can optimize the Hamiltonian of L0RBCS with respect to $q$ under the condition that $r$ is fixed, whereas Algorithm \ref{alg5} (Jacobi SOR) can optimize the Hamiltonian of L0RBCS with respect to $r$ under the condition that $q$ is fixed.

As shown in Fig. \ref{fig:algorithm_relation}, the architecture we have proposed alternate repeatedly between open-loop CIM or closed-loop CIM, and Jacobi SOR to perform alternative optimization of the Hamiltonian in Eq. \ref{eq.L0-regularization-based-CS-Hamiltonian} \cite{RN1141,RN1142,RN1204}. Details of the alternative optimization are shown in Algorithm \ref{alg6}. As mentioned above, $\lambda$ in Eq. \ref{eq.L0-regularization-based-CS-Hamiltonian} is related to $\eta$ in Eq. \ref{eq.2} by $\eta=\sqrt{2\lambda}$ \cite{RN1141}. $\lambda$ is set to satisfy this relationship.

\begin{algorithm}
  \caption{Alternating minimization of L0RBCS as a QUBO problem. The schedules of the pump rate and threshold are given in Section 2.7}
  \label{alg6}
\begin{algorithmic}[1]
\REQUIRE J matrix: $J = -A^T A + \sum_k\gamma_k \Gamma_k^T\Gamma_k$, Zeeman terms: $g = A^T y$, Initial value: $r_{init}$, Pump rate sequence: $p\in\Re^{N_{step}}$, Threshold sequence: $H\in\Re^{N_{outer}}$
 \ENSURE $N$-dimensional support vector: $q$, $N$-dimensional signal vector: $r$
 \STATE $r=r_{init}$.
 \FOR{n=1 to $N_{outer}$}
 \STATE $\eta \leftarrow H_n$ (if open-loop) or $\lambda \leftarrow H_n^2/2$ (if closed-loop). 
 \STATE Minimize $\mathcal{H}$ with respect to $\sigma$ by QUBO-open-loop CIM (Algorithm \ref{alg1}) or QUBO-closed-loop CIM (Algorithm \ref{alg3}).
 \STATE Minimize $\mathcal{H}$ with respect to $r$ by Jacobi SOR (Algorithm \ref{alg5}).
 \ENDFOR
 \RETURN $\sigma$ and $r$
\end{algorithmic}
\end{algorithm}

\subsection{Parameter configuration}

Parameter settings for the experiments comparing FPGA and GPU implementations are described below. Parameter values for each algorithm on each experiment are summarized in Table \ref{table:Parameter}. These values were determined based on our previous research results \cite{RN1144,RN1141,RN1142,RN1204}. The schedules of the pump rate $p$ and threshold $\eta$ were given as the following functions, and parameter values of these functions are also shown in Table \ref{table:Parameter}.

The schedule of the pump rate $p$ for open-loop CIM (Algorithm \ref{alg1}) was set as follows.
\begin{eqnarray}
p(t) = p_{max} \left(\frac{t}{N_{step} \Delta t}\right)^2, \nonumber
\end{eqnarray}
where $t$ is the time in Eq. \ref{eq.1}, normalized to the photon lifetime. $N_{step}$ and $\Delta t$ are the number of loop repetitions and time interval for the TE calculation in Algorithm \ref{alg1}, and $N_{step} \Delta t$ corresponds to the end time of TE. $p_{max}$ is the maximum value of the pump rate. These parameter settings are shown in Table \ref{table:Parameter}.

The schedule of the pump rate $p$ for closed-loop CIM (Algorithm \ref{alg3}) was given by the following.
\begin{eqnarray}
p(t)= p_{tr} - dp + \frac{2 dp}{1+\exp\left(-\left(t-4\right)/2\right)}. \nonumber
\end{eqnarray}
Here, $t$ is the time in Eq. \ref{eq.4}, normalized to the photon lifetime, and $p_{tr}$ is the baseline of the pump rate, and $dp$ is the variation range of the pump rate. These parameter settings are shown in Table \ref{table:Parameter}.

The schedule of the threshold $\eta$ in the alternating optimization for L0RBCS (Algorithm \ref{alg6}) is given by the following function for both open-loop CIM and closed-loop CIM.
\begin{eqnarray}
\eta(n) = \max\left(\eta_{init}\frac{n-1}{N_{outer}-1},\eta_{end}\right), \nonumber
\end{eqnarray}
where $n$ is the loop step for alternating optimization (a natural number), and $N_{outer}$ is the number of loop repetitions. $\eta_{init}$ is the initial value for the threshold, and $\eta_{end}$ is the final value for the threshold. The settings for these parameters for each experiment are shown in Table \ref{table:Parameter}. 

\begin{table*}[t]
 \caption{Parameter settings for each experiment}
 \label{table:Parameter}
 \centering
  \begin{tabular}{lllll}
   \hline
     Fig. $\#$ & Open-loop CIM (Alg. \ref{alg1}) & Closed-loop CIM (Alg. \ref{alg3}) & Jacobi SOR (Alg. \ref{alg5}) & Alternating mini. (Alg. \ref{alg6})\\
     \hline
     \hline
     Figs. 6 $\&$ 7 & $\Delta t=0.1$, $p_{max}=2$, $\chi = {\rm identity}$ & $\Delta t=0.02$, $p_{tr}=1$, $dp=0.6$ & - & -\\
     & $K=0.5$, $N_{step}=101$, $g_s^2=10^{-7}$  & $\beta=1$, $\tau=1$, $K=0.1$, $N_{step}=501$ &  &\\
     \hline
     Fig. 8 & $\Delta t=0.1$, $p_{max}=1.5$, $\chi = {\rm absolute}$ & $\Delta t=0.02$, $p_{tr}=1$, $dp=0.4$  & $\Delta t=0.3$ & $N_{outer}=51$\\
     & $K=0.25$, $N_{step}=51$, $g_s^2=10^{-7}$ & $\beta=1$, $\tau=1$, $K=0.1$, $N_{step}=1001$ &  $N_{step}=1001$ & $\eta_{init}=0.8$, $\eta_{end}=0.18$\\
     \hline
     Fig. 9 & $\Delta t=0.1$, $p_{max}=1.5$, $\chi = {\rm absolute}$ & $\Delta t=0.02$, $p_{tr}=1$, $dp=0.6$  & $\Delta t=0.1$ & $N_{outer}=11$\\
     & $K=0.25$, $N_{step}=51$, $g_s^2=10^{-7}$ & $\beta=1$, $\tau=1$, $K=0.1$, $N_{step}=501$ & $N_{step}=1001$ & $\eta_{init}=\eta_{end}$ (see main text)\\
     \hline
  \end{tabular}
\end{table*}

\begin{figure}[htbp]
\includegraphics[width=100mm]{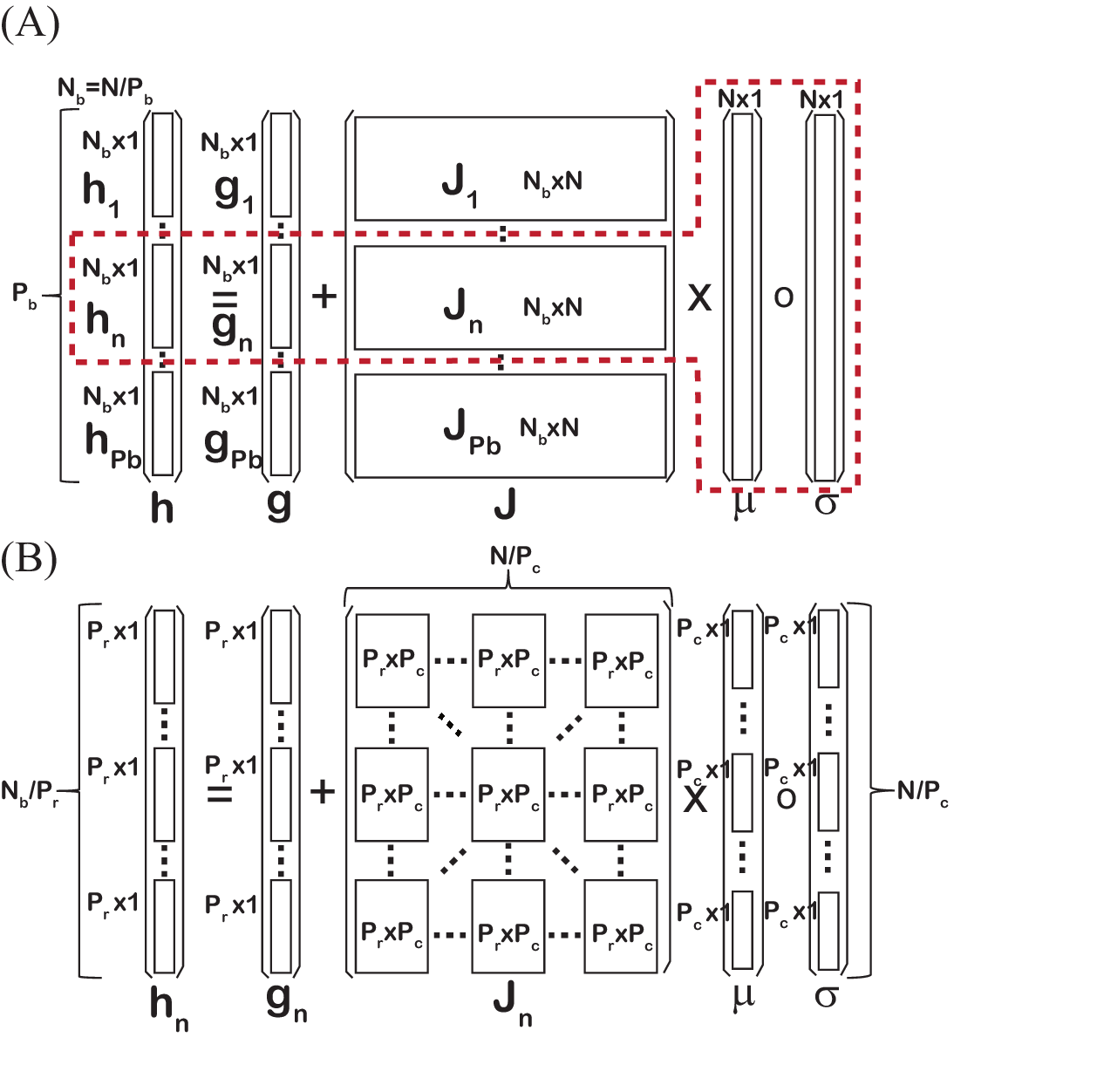}
\caption{Indexes $P_b$, $P_r$, and $P_c$, showing the parallelization format used for matrix-vector multiplication in the local field calculation.
(A) Block parallelization index $P_b$ of local field calculation. The local field calculation is partitioned into $P_b$ blocks, and the process in (B) is applied to each block in parallel.
(B) Matrix-vector multiplication parallelization indices $P_r$ and $P_c$ of block-partitioned local field calculation. In a parallel MAC operation, a $P_r\times P_c$ matrix and $P_c$-dimensional vector are multiplied. By repeating this $N/P_c$ times, $P_r$ entries of local field have been calculated. This local calculation is repeated $N_b/P_r$ times to have completely calculated all entries of local field. $N_b=N/P_b$.}
\label{fig.matrix_para}
\end{figure}

\begin{figure*}[htbp]
\begin{center}
  \includegraphics[width=160mm]{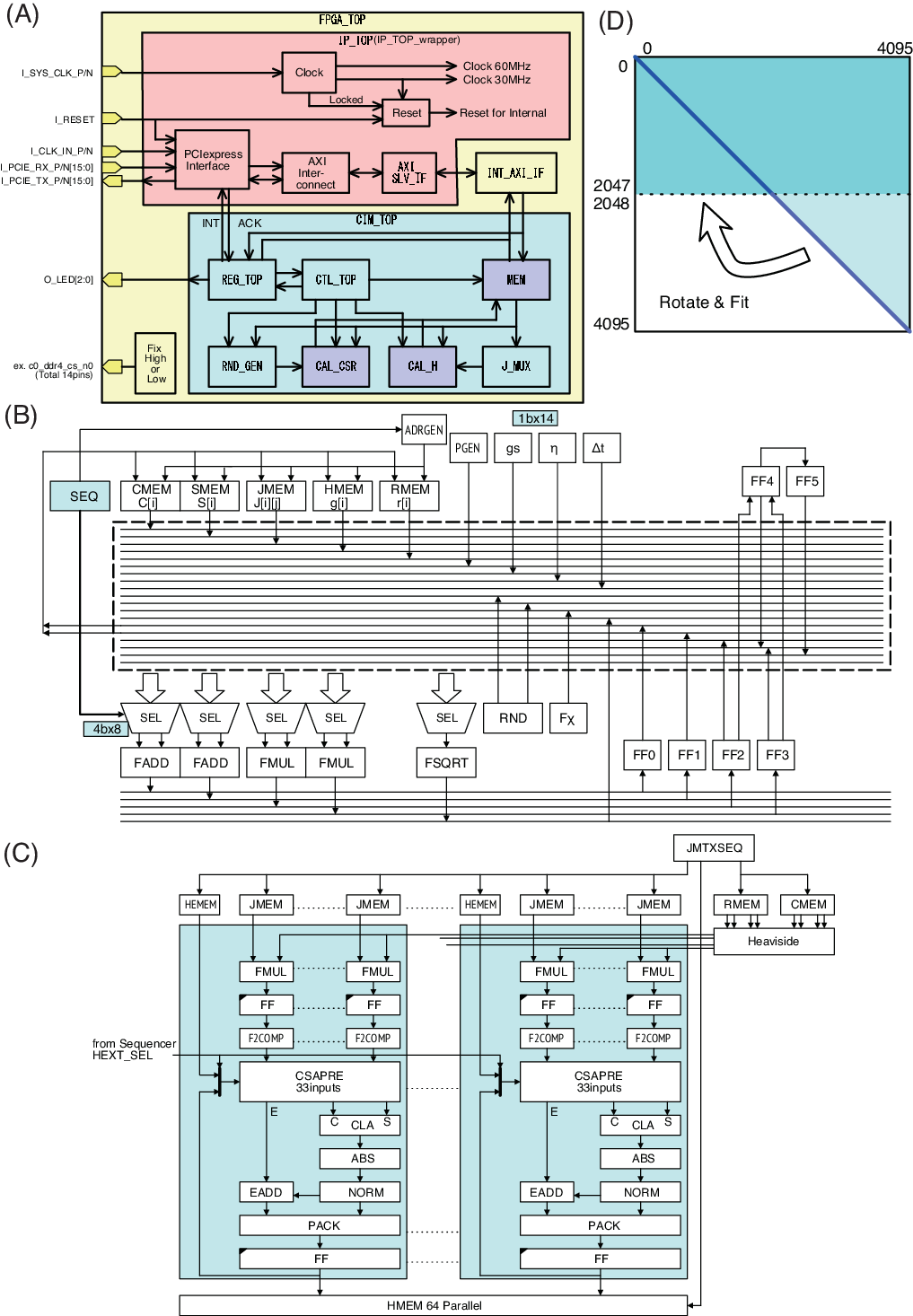}
\end{center}
\caption{FPGA architecture.
(A) Block diagram showing the relationships between individual FPGA modules. An overview of each module is given in Table \ref{table:FPGA_module_layer}.
(B) CAL$\_$CSR functional diagram.
(C) CAL$\_$H functional diagram.
(D) Storage scheme for matrix $J$. }
\label{fig.FPGA_arc}
\end{figure*}

\begin{table*}[t]
 \caption{Hierarchy and overview of implemented modules}
 \label{table:FPGA_module_layer}
 \centering
  \begin{tabular}{lllll}
   \hline
     Layer 1 & Layer 2 & Layer 3 & Layer 4 & Overview \\
   \hline \hline
   FPGA$\_$TOP & & & & Whole circuit for CIM \\
     & IP$\_$TOP & & & External interface and clock generator\\
     &  & IP cores & & PCIe interface, Phase locked loop, AXI interconnect, etc. \\
     &  & AXI$\_$SLV$\_$IF & & AXI slave interface coded in RTL\\
   & CIM$\_$TOP & & & Main circuit for CIM coded in RTL\\
   & & REG$\_$TOP & & Register sets for storing parameters and settings\\
   & & CTL$\_$TOP & & Module for controlling calculation and memory access\\
   & & CAL$\_$CSR & & Modules for time evolution calculation. 64 parallel modules\\
   & & CAL$\_$H & & MAC modules for local field calculation. 64 parallel modules\\
   & & MEM & & Memory group for storing data\\
   & & & JMEM & Store J matrix\\
   & & & HMEM & Store local field $h$\\
   & & & HEMEM & Store Zeeman terms $g$\\
   & & & CMEM & Store in-phase amplitudes $c$\\
   & & & SMEM & Store quadrature phase amplitudes $s$ or feedback errors $e$\\
   & & & RMEM & Store auxiliary variables $r$\\
   & & & ADMEM & Store inverse of J's diagonal elements $d$\\
   & & & PMEM & Store pump rate sequence $p$\\
   & & RND$\_$GEN & & Normal random number generator\\
   & & J$\_$MUX & & Read out symmetric matrix from stored upper triangular matrix data\\
   & INT$\_$AXI$\_$IF & & & AXI bus interface \\
   \hline
  \end{tabular}
\end{table*}

\section{Experimental environment}

We evaluate the computation accuracy and running time of Algorithms 1 to 4 on the following FPGA and GPU systems.

For the GPU implementation, we used an NVIDIA Quadro RTX 8000 (4608 CUDA cores, 48 GB GDDR6 memory, 16.3 TFLOP 32-bit floating-point operations). We used a machine with an AMD RIZEN9 4950X CPU and 128GB of memory, connected to the GPU by 16 PCI Express 3.0 interfaces. The OS was Ubuntu 22.04.3. Algorithms 1 to 4 were programmed in CUDA. The CUDA programs were compiled using the mexcuda compile command in MATLAB, and linked to shared libraries called MEX functions that can be called from MATLAB. In MATLAB, we performed pre-processing such as generating the $J$ matrix and the Zeeman term. These were passed to the MEX functions to perform calculation on the GPU, and the results were returned to MATLAB through the MEX functions, for graphing and other post processing.

For the FPGA implementation, we used Xilinx ALVEO U250s (INT8 TOPs (peak) 33.3, 1,728K look-up tables (LUTs), 54MB internal SRAM, 38Tb/s total internal SRAM bandwidth). We used a machine with an AMD RIZEN9 3950X CPU and 64GB of memory, which was connected to the FPGA by 16 PCI Express 3.0 interfaces. The OS was Ubuntu 22.04.3. We developed an application programming interface (API) to control the FPGA. This API was used to program Algorithms 1 to 4 in the C programming language. These coded C programs were compiled and linked into MEX functions and static libraries of the FPGA control APIs. As with the GPU case, we performed preprocessing in MATLAB as described above, passed the data to MEX functions to be processed by the FPGA, and the results from the FPGA were passed back to MATLAB through the MEX functions for graphing and other post processing.

\section{FPGA architecture}

\subsection{Parallelization scheme and index definitions}

In this section, we redefine the indices $P_b$, $P_r$ and $P_c$, introduced to explain the parallelization scheme in Toshiba's FPGA implementation of SB (FPGA-SB) \cite{RN1139,RN1135}.

Fig. \ref{fig.matrix_para} shows the parallelization scheme for matrix-vector multiplication in the local field calculation that is assumed for these indices. This schematic diagram shows the parallelization into $P_b$ blocks (Fig. \ref{fig.matrix_para} (A)) and the $P_r\times P_c$ parallel multiply–accumulate (MAC) operations in each of $P_b$ blocks (Fig. \ref{fig.matrix_para} (B)). The local field calculation is firstly partitioned into $P_b$ blocks as shown in Fig. \ref{fig.matrix_para} (A). The local field $h$, Zeeman term $g$ and coupling matrix $J$ are allocated for the $P_b$ blocks to get $\bf h_n$, $\bf g_n$ and $\bf J_n$ $(n=1,\cdots,P_b)$. For each of the partitioned blocks, the following equation holds.
\begin{eqnarray}
{\bf h_n}={\bf g_n}+{\bf J_n} \left(\mu \circ \sigma\right), \ \ n=1,\cdots,P_b. \nonumber
\end{eqnarray}
These calculations are independent of each other, so they can be operated in parallel for each block. For each of these $P_b$ blocks, the following process is performed in parallel.

As shown in Fig. \ref{fig.matrix_para} (B), the matrix-vector multiplication is performed for each block as follows. There are $P_r$ modules of $P_c$-input multiplier-accumulator (MAC), which are used to operate the multiplication of $P_r\times P_c$ submatrices of $\bf J_n$, with $P_c$-dimensional partial vectors of $\mu \circ \sigma$, in parallel. This is repeated $N/P_c$ times to complete the calculation of $P_r$ entries of ${\bf h_n}$. Then, this calculation is repeated $N/P_b/P_r$ times to complete the calculation of ${\bf h_n}$. These operations are performed on $P_b$ blocks, in parallel.

There are also assumed to be $P_b\times P_r$ parallel modules to perform the TE calculation, which is the same as the $P_b\times P_r$ parallelism as the local field calculation described above.

\subsection{Architecture overview}

The implementation we present is highly versatile and capable of performing the three algorithms above on an identical architecture, as well as being able to represent the coupling matrix $J$ with FP32 accuracy up to the size of $4096\times 4096$. We now give an overview of the architecture we have developed to achieve this. Fig. \ref{fig.FPGA_arc} (A) is a block diagram showing the relationships among each of the modules implemented in the FPGA. The modules in the red part (IP$\_$TOP) include the external interfaces and clock generation implemented using mainly Xilinx intellectual property (IP) core functions. The modules in the other parts (CIM$\_$TOP and INT$\_$AXI$\_$IF) mainly compute the algorithms described above, which are coded in register transfer level (RTL). Table \ref{table:FPGA_module_layer} gives the hierarchy and overview of these modules shown in Fig. \ref{fig.FPGA_arc} (A). CTL$\_$TOP, CAL$\_$H, CAL$\_$CSR and J$\_$MUX, which perform the algorithm computations, are described below.

The calculation control module CTL$\_$TOP controls the calculation sequence. 
For the FPGA calculation, the MAC process in CAL$\_$H, the multipliers and accumulators in CAL$\_$CSR and so on are switched for each cycle. Settings to memory addresses and the switching configuration for multiplexers for each cycle are stored in control code memory.
Then, each stored content is sequentially read from the memory at each cycle to control the calculation sequence. Therefore, by rewriting the control code on the memory, various algorithms such as not only the CIM algorithms but also the SB algorithm can be calculated in our architecture in principle.

The functional diagram of the CAL$\_$CSR module, which calculate TE of each algorithm, is shown in Fig. \ref{fig.FPGA_arc} (B). Memory (in MEM) also appears in the schematic diagram to show connections clearly, but this memory is not actually included in the CAL$\_$CSR module. Details of MEM are given in Table \ref{table:FPGA_module_layer}. The values of the various parameters ($g_s$, $\eta$, $\tau$, $\beta$, $\Delta t$, etc.) are read out from registers where they are stored. The pump rate for each step is read from PGEN in MEM. CAL$\_$CSR is composed of two accumulators (FADD), two multipliers (FMUL), a square-root module (FSQRT), a module to perform $F_\xi$ described above and six flip-flops (FFs) to temporarily store calculation results, and these are connected to a databus through selector (SEL). A random-number generator (RND) is also connected to the databus. All these arithmetic modules are FP32. The calculation proceeds by switching the inputs and outputs of these arithmetic modules and the FFs at each cycle under the sequential control from the CTL$\_$TO described above. The CAL$\_$CSR has 64 branches in our FPGA system, so it can calculate 64 TE equations simultaneously. 

A functional diagram of CAL$\_$H, which calculate the local field, is shown in Fig. \ref{fig.FPGA_arc} (C). Memory (in MEM) also appears in the schematic diagram to show connections clearly, but this memory is not actually included in the CAL$\_$H module. Details of MEM are shown in Table \ref{table:FPGA_module_layer}. Heaviside in Fig. \ref{fig.FPGA_arc} (C) switches between $\mu\leftarrow c$ for Ising optimization, and $\sigma \leftarrow H(c)$ for QUBO, and performs the element-wise multiplication $\mu \circ \sigma$. Then inner products of 32-element vectors are performed in one cycle by 32-input FP32 MACs. The CAL$\_$H is composed of 64 parallel MAC modules in our FPGA system, so it can multiply a $64\times 32$ matrix with a 32-element vector in one cycle. Therefore, there are 2048 MAC processing elements (PEs). Under the sequential control from CTL$\_$TO, CAL$\_$H also performs addition of the Zeeman term (stored in HEMEM) to the local field, as shown in Fig. \ref{fig.FPGA_arc} (C). However, to prioritize parallel processing, CAL$\_$H does not conform to rounding rules in the IEEE 754 FP32.

 It is difficult to store the $4096\times 4096$ coupling matrix $J$ in FP32 in the internal memory of the FPGA we are using (Xilinx ALVEO U250) while reserving space for the other variables and parameters.
 However, using external memory would greatly reduce the speed of the calculation. Since almost all optimization problems we deal with have symmetric coupling matrices, it is sufficient to store the upper triangle of coupling matrix $J$, so only half the space needed for the entire matrix needs to be stored. As such, we store the matrix $J$ in the internal memory in a special format shown in Fig. \ref{fig.FPGA_arc} (D). However, for the calculation of CAL$\_$H it is necessary to read the original symmetric matrix from the $J$ data stored in this special format. J$\_$MUX is a multiplexer that fills-in the symmetric matrix entries from the data in memory. This enables us to implement the local field calculation under full-coupling with FP32 representation up to the system size of $N=4096$ on a single FPGA.

Table \ref{table:FPGA_CIM} summarizes the specifications of the FPGA architecture we have constructed. For comparison, the specifications for FPGA-SB are shown in Table \ref{table:FPGA_SB} \cite{RN1139,RN1135}. Our architecture allows the configuration of Zeeman terms with FP32, which is not configurable in the FPGA-SB. In the FPGA-SB, the entries of the coupling matrix $J$ are represented with one bit as $[-1,1]$, but in our implementation, those of $J$ are represented with FP32 values. For calculating TE of equations, FPGA-SB uses 16-bit fixed-point operations, while our system uses FP32 operations.

For the parallelization indices for the local field calculation defined in the previous section, FPGA-SB uses $P_r=32$, $P_c=32$ and $P_b=8$, while our implementation uses $P_r=64$, $P_c=32$ and $P_b=1$. Thus, the number of MAC PEs in our implementation is 2048 while that in FPGA-SB is 8192, which is four times as many as our implementation. However, in our implementation the coupling marix $J$ is represented with FP32, while in FPGA-SB that is represented with one bit. Table \ref{table:FPGA_CIM} also summarizes the number of cycles and processing time per TE step when processing Algorithms \ref{alg1}, \ref{alg3} and \ref{alg5} for different system sizes $N$. Detail evaluations of the numbers of cycles and processing times are given in the next sections.

\begin{table}[h]
 \caption{Specification of FPGA Cyber CIM}
 \label{table:FPGA_CIM}
 \centering
  \begin{tabular}{llll}
   \hline
     & $N=1024$ & $N=2048$ & $N=4096$ \\
   \hline \hline
   {\bf Zeeman terms} & Configurable & Configurable & Configurable \\
   \hline
   {\bf Architecture} & & & \\
   Precision of J & Single float & Single float & Single float \\
   Precision of TE & Single float & Single float & Single float\\
   $P_r/P_c/P_b$ & 64/32/1 & 64/32/1 & 64/32/1 \\
   $\#$ of MAC PEs & 2048 & 2048& 2048 \\
   \hline
   {\bf Cycles per step} & & & \\
   Algorithm 1 & 1030 & 3075 & 10245 \\
   Algorithm 2 & 1030 & 3075 & 10245 \\
   Algorithm 3 & 580 & 2180 & 8453 \\
   \hline
   {\bf Processing time per step [$\mu s$]} & & & \\
   Algorithm 1 & 34.3 & 102.5 & 341.5 \\
   Algorithm 2 & 34.3 & 102.5 & 341.5 \\
   Algorithm 3 & 19.3 & 72.7 & 281.8 \\
\hline
{\bf Operating Clock Frequency} & & & \\
$F_{sys}$[Mhz] & 30 & 30 & 30 \\
   \hline
  \end{tabular}
\end{table}

\begin{table}[h]
 \caption{Specification of FPGA SB \cite{RN1139,RN1135}}
 \label{table:FPGA_SB}
 \centering
  \begin{tabular}{llll}
   \hline
     & $N=2048$ & $N=4096$ \\
   \hline \hline
   {\bf Zeeman terms} & none & none \\
   \hline
   {\bf Architecture} & & \\
   Precision of J & 1 bit & 1 bit \\
   Precision of TE & 16-bit fixed point & 16-bit fixed point \\
   $P_r/P_c/P_b$ & 32/32/8 & 32/32/8 \\
   $\#$ of MAC PEs & 8192 & 8192 \\
   Cycles per step & 624 & 2224  \\
   Processing time per step [$\mu s$] & 2.2 & 8.3 \\
\hline
{\bf Operating Clock Frequency} & & \\
$F_{sys}$[Mhz] & 279 & 269 \\
   \hline
  \end{tabular}
\end{table}

\begin{table}[h]
 \caption{Estimated cycles per step}
 \label{table:EST_CYC}
 \centering
  \begin{tabular}{lllll}
   \hline
     & $C_e$ & $N=1024$ & $N=2048$ & $N=4096$ \\
   \hline
   Algorithm 1 & 32 & 1024 & 3072 & 10240 \\
   Algorithm 2 & 32 & 1024 & 3072 & 10240 \\
   Algorithm 3 & 4 & 576 & 2176 & 8448 \\
\hline
   SB & ? & - & 576 & 2176 \\
   \hline
  \end{tabular}
\end{table}

\subsection{Evaluation of number of cycles}

In this section, according to the indices of the parallelization scheme described above, we evaluate the number of cycles required by the proposed architecture to perform calculations.

\begin{figure}[htbp]
\includegraphics[width=90mm]{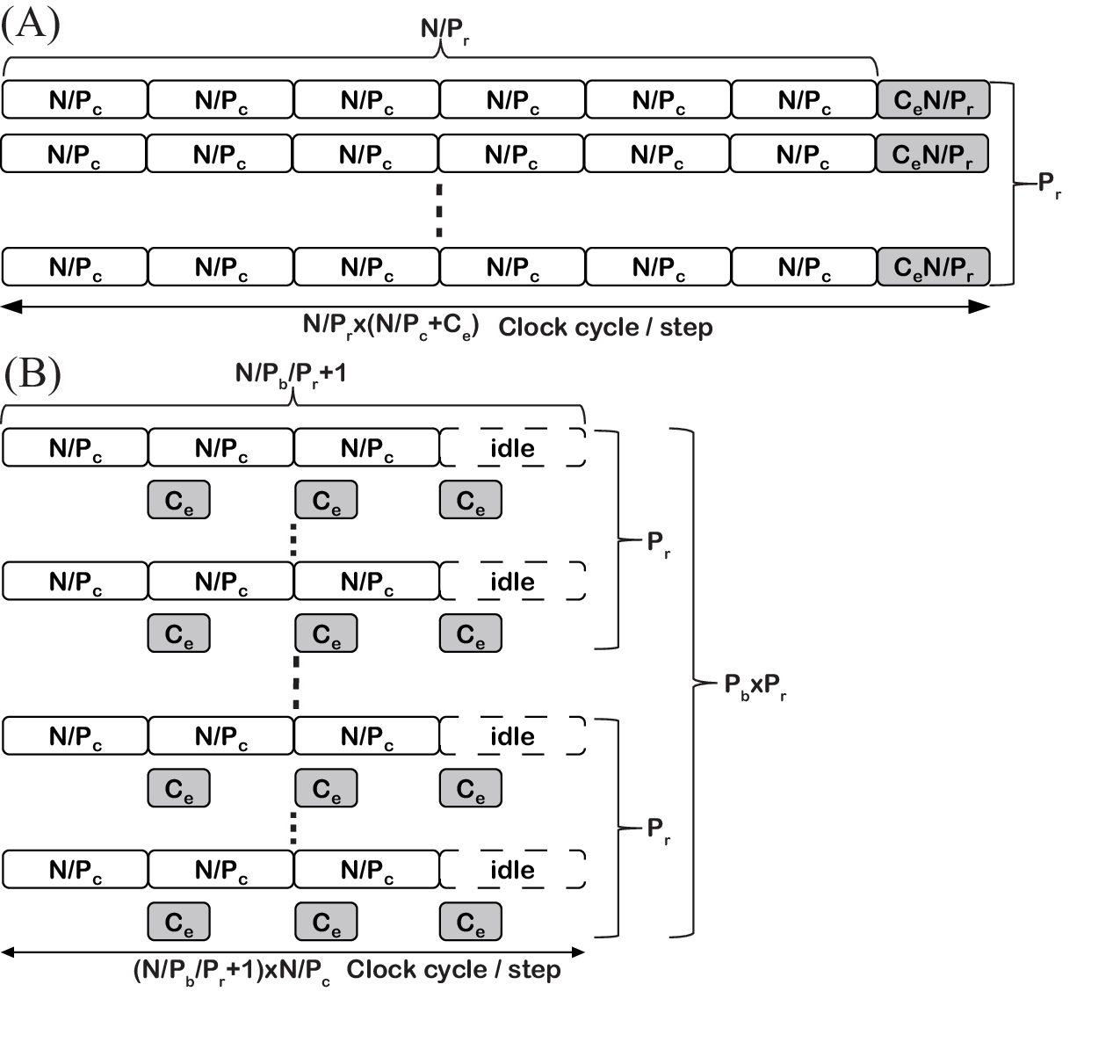}
\caption{Parallelization scheme and number of cycles per step.
(A) Our architecture ($P_b=1$).
(B) FPGA-SB ($P_b>1$).}
\label{fig.timechart_para}
\end{figure}

We first evaluate the number of cycles needed to process one step including the local field and TE calculations for Algorithms \ref{alg1}, \ref{alg3} and \ref{alg5}. Fig. \ref{fig.timechart_para} (A) shows a schematic diagram of the number of cycles required for one step in the parallelization scheme used by the proposed architecture. Our FPGA implementation does not adopt the block parallelization, so $P_b=1$. For processing of one step, the TE computation is performed immediately after the local field calculation. The individual MAC operation for local field calculation requires one cycle and the individual TE calculation requires different numbers of cycles depending on the algorithm. To derive a general expression that does not depend on the algorithm implemented, we express the number of cycles for TE as $C_e$. As described above, a $P_r\times P_c$ parallel MAC operation is repeated $N/P_c$ times, and this calculation is repeated $N/P_r$ times to complete all local field calculations. Thus, the calculation of local field requires $N/P_c\times N/P_r$ cycles. For the TE calculation, there are $P_r$ modules so calculating $N$ TE equations will require $N/Pr\times C_e$ cycles. Totaling these cycle counts gives an estimate of the number of cycles required to process one step.
\begin{eqnarray}
N/P_r\times \left(N/P_c + C_e\right). \label{eq.FPGA_CIM_cycle}
\end{eqnarray}
As shown in Table \ref{table:FPGA_CIM}, our FPGA architecture adopts $P_b=1$, $P_r=64$ and $P_c=32$. Table \ref{table:EST_CYC} gives $C_e$ for each Algorithm and estimates of the number of cycles per step depending on the system size $N$. There is a difference of several cycles between these estimates and the measured values shown in Table \ref{table:FPGA_CIM}. These differences are due to processing overhead.

\subsection{Comparison with cycles for FPGA-SB}

We now compare the number of cycles required for FPGA-SB with that required for our system.

FPGA-SB adopts block parallelization, so $P_b>1$. Fig. \ref{fig.timechart_para} (B) shows a schematic diagram of the number of cycles per step in the parallelization scheme used by  FPGA-SB \cite{RN1139}. The local field and TE calculations are parallelly operated by shifting those initial timings as shown in Fig. \ref{fig.timechart_para} (B). The individual MAC operation for local field calculation requires one cycle and the individual TE calculation requires $C_e$ cycles. As described above, the $P_r\times P_c$ parallel MAC operation is repeated $N/P_c$ times, and this calculation is repeated $N/P_b/P_r$ times. These computations are performed on $P_b$ blocks in parallel, to complete all local field calculations. As such, the number of cycles required for local field calculations is $N/P_c\times N/P_b/P_r$. Then, as shown in Fig. \ref{fig.timechart_para} (B), the TE calculation is executed after the first $N/P_c$ cycles in which the local field calculation required for this calculation is completed, and thereafter similarly executed every $N/P_c$ cycles. In order to secure the interval necessary for the final TE calculation, an idle interval of $N/P_c$ cycles is reserved. Thus, an estimate of the number of cycles required to process one step is given by:
\begin{eqnarray}
\left(N/P_b/P_r + 1\right) \times N/P_c
\end{eqnarray}
As shown in Table \ref{table:FPGA_SB}, the single-FPGA implementation of SB adopts $P_b=8$, $P_r=32$ and $P_c=32$ \cite{RN1139,RN1135}. Table \ref{table:EST_CYC} shows the estimated number of cycles for FPGA-SB depending on the system size $N$. These estimates differ from the measured values in Table \ref{table:FPGA_SB} by about 50 cycles. These differences are due to processing overhead. 

\begin{figure}[htbp]
\begin{center}
  \includegraphics[width=70mm]{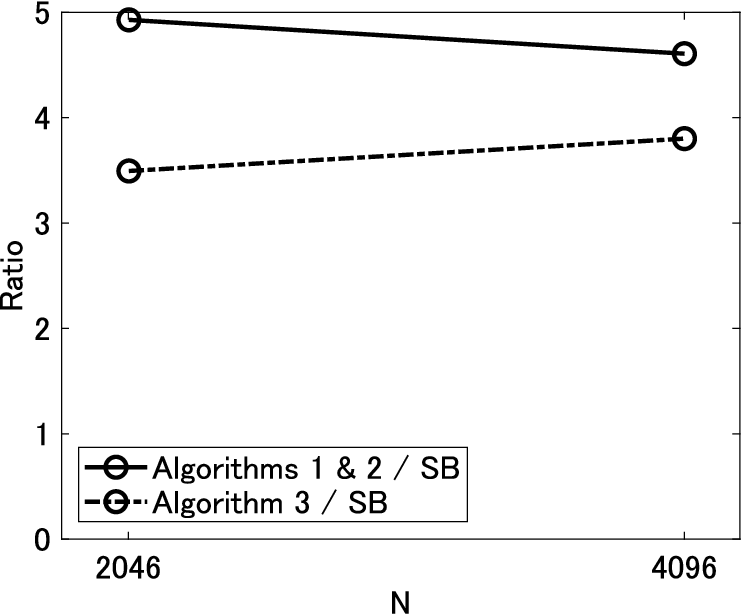}
\end{center}
\caption{Ratios of cycles-per-step for our architecture and FPGA-SB for $N=2048$ and $N=4096$ }
\label{fig.cycle_ratio}
\end{figure}

As can be seen in Fig. \ref{fig.timechart_para}, the main reason of the difference in the number of cycles per one step between our FPGA system and FPGA-SB is the difference in the degree of parallelism. The number of parallels in our architecture is $P_r=64$, while that in FPGA-SB is $P_b \times P_r=256$, so we can estimate that our architecture requires about four-times as many cycles as FPGA-SB. We calculate the ratios of number of cycles per step for our architecture (Table \ref{table:FPGA_CIM}) and FPGA-SB (Table \ref{table:FPGA_SB}), and show these in Table \ref{table:cycle_ratio} and Fig. \ref{fig.cycle_ratio}. For Algorithms 1 and 2, the ratio is greater than four, and for Algorithm 3, it is less than four. This difference is due to the difference in the size of $C_e$. However, each entry of the coupling matrix $J$ in our architecture is a single-precision floating point representation, whereas each entry of $J$ in FPGA-SB is a 1 bit representation. 

\begin{table}[h]
 \caption{Ratios of cycles per step for our architecture and FPGA-SB}
 \label{table:cycle_ratio}
 \centering
  \begin{tabular}{lll}
   \hline
     & $N=2048$ & $N=4096$ \\
   \hline
   Algorithm 1 / SB & 4.928 & 4.607 \\
   Algorithm 2 / SB & 4.928 & 4.607 \\
   Algorithm 3 / SB & 3.494 & 3.801 \\
   \hline
  \end{tabular}
\end{table}

\section{Performance evaluation}

Here, we evaluate the performance of the FPGA system we have developed. As evaluation benchmarks, we used the CDMA multi-user detector, which is an optimization problem in the form of the Ising Hamiltonian including Zeeman terms, and L0RBCS, which is an optimization problem in the form of the QUBO Hamiltonian including Zeeman terms. Earlier large-scale FPGA systems implementing such as SB and CIM have not been able to handle these problems \cite{RN1050,RN1139,RN1135,RN1192,RN1133}. Below, we evaluate the performance of our FPGA system for these problems through comparing calculated results and running times with those obtained using the GPU described in the section {\bf Experimental environment}. 

\subsection{CDMA multi-user detector (Ising Hamiltonian)}

This section focuses on the CDMA multi-user detector, which is an optimization problem in the form of the Ising Hamiltonian with Zeeman terms \cite{RN1144}. In the section {\bf Implemented Algorithms}, we described the optimization problem for the CDMA multi-user detector used in this research. Here, for each entry of the spreading code series, $\xi_i^\mu$, is set to either $+1$ or $-1$ independently with the probability of $1/2$ with respect to $i$ and $\mu$. In this experiment, each user’s bit of the information data bits, $\sigma_i$ is set to $1$. Note that an arbitrary bit string of $[\sigma_1,\cdots,\sigma_N]$ can be transformed to $[1,\cdots,1]$ by performing a Gauge transformation (i.e. a variable transformation with $\sigma_i c_i:=c_i$). We then compared the calculated results and running times for parallel implementations with FPGA and GPU of both Algorithm \ref{alg1} (open-loop CIM) and Algorithm \ref{alg3} (closed-loop CIM).

\begin{figure}[htbp]
\includegraphics[width=90mm]{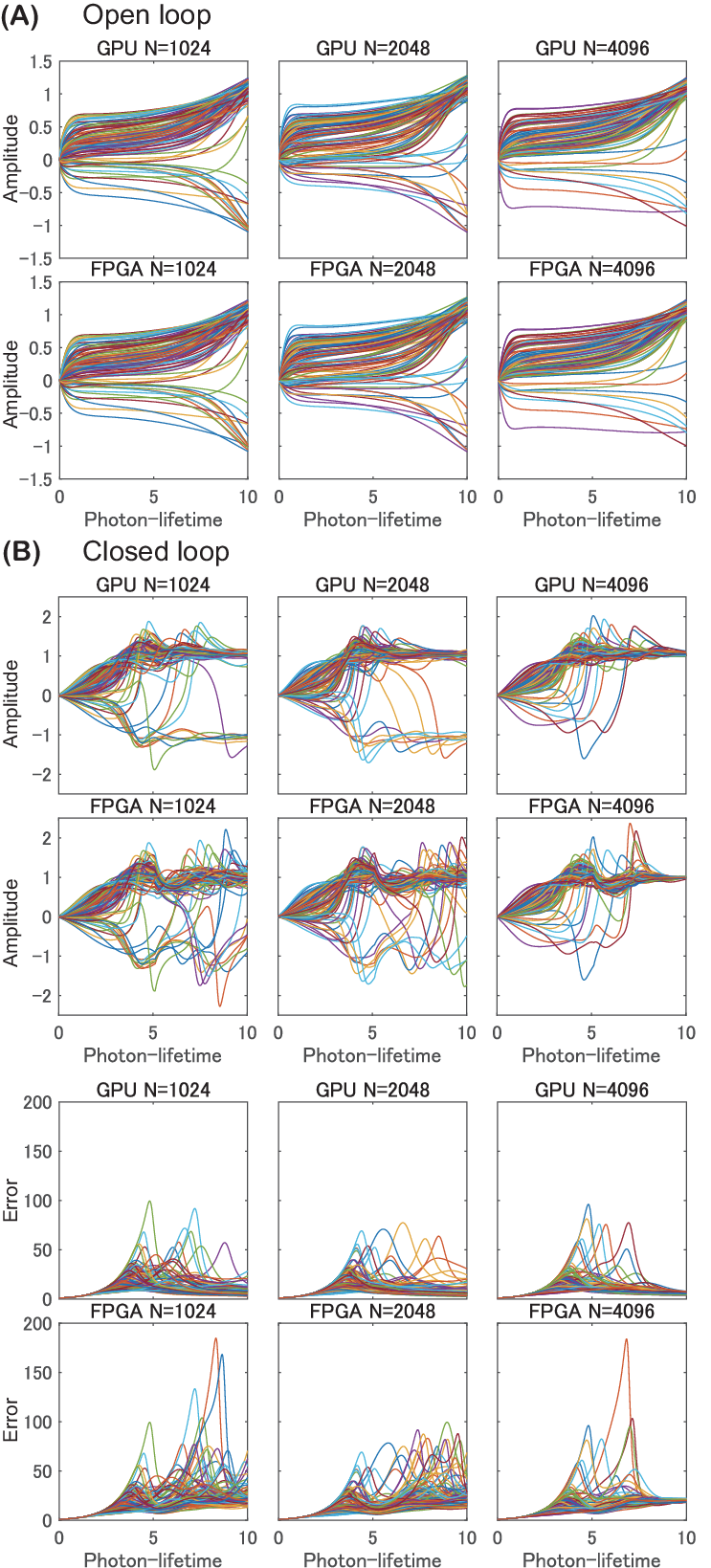}
    \caption{Trajectories during solution exploration by cyber CIMs on GPU and FPGA. Application of the CDMA multi-user detector. (A) Open-loop CIM. Temporal profiles of OPO pulse amplitudes $c_1,\cdots,c_{100}$. (B) Closed-loop CIM. Temporal profiles of OPO pulse amplitudes $c_1,\cdots,c_{100}$ and feedback error $e_1,\cdots,e_{100}$. For both (A) and (B), Spreading code series and initial values used in cyber CIMs on GPU and FPGA were the same. The system sizes of these are $N=1024$, $N=2046$ and $N=4096$. All $\alpha=0.6$. }
\label{fig.cdma_orbit}
\end{figure}

\subsubsection{Comparison of trajectories during solution exploration}

Examples of trajectories during solution exploration by open-loop CIM and closed-loop CIM on GPU and FPGA are shown in Fig. \ref{fig.cdma_orbit}. In this experiment, using the spreading code series randomly generated with the same random seed under the same diffusion rate of $\alpha=0.6$, the same coupling matrix $J$ and Zeeman terms were configured for all CIMs on GPU and FPGA. For open-loop CIM on both GPU and FPGA in Fig. \ref{fig.cdma_orbit} (A), the initial values for OPO pulse amplitudes were the same, with $c_{init}=0$ and $s_{init}=0$. For closed-loop CIM on both GPU and FPGA in Fig. \ref{fig.cdma_orbit} (B), the initial values for OPO pulse amplitudes were the same, with $c_{init}$ set to the same random values from a standard normal distribution with a mean of 0 and a variance of 0.02, and initial values of feedback errors were also the same, with $e_{init}=1$. Parameter values for this experiment are shown in Table \ref{table:Parameter}. Three system sizes were used: $N=1024$, $N=2046$ and $N=4096$.

Figure \ref{fig.cdma_orbit} (A) shows 100 OPO pulse amplitudes, $c_1,\cdots,c_{100}$ during solution exploration by open-loop CIM on GPU and FPGA. Figure \ref{fig.cdma_orbit} (B) shows 100 OPO pulse amplitudes, $c_1,\cdots,c_{100}$, and 100 feedback errors, $e_1,\cdots,e_{100}$ during solution exploration by closed-loop CIM on GPU and FPGA. Under the above conditions, open-loop CIMs on GPU and FPGA had the same solution trajectories for each system size $N$. In contrast, even though closed-loop CIMs on GPU and FPGA had the same conditions including initial values, those solution trajectories differed. This difference in solution trajectories for closed-loop CIM may have been because our FPGA system does not adhere to rounding rules of IEEE 754 FP32. Closed-loop CIM has more dynamical complexity due to the CAC \cite{RN1159}, so it is more sensitive to this sort of error from rounding in MAC, which could account for differences in solution trajectory for GPU and FPGA.

\begin{figure*}[htbp]
\begin{center}
  \includegraphics[width=150mm]{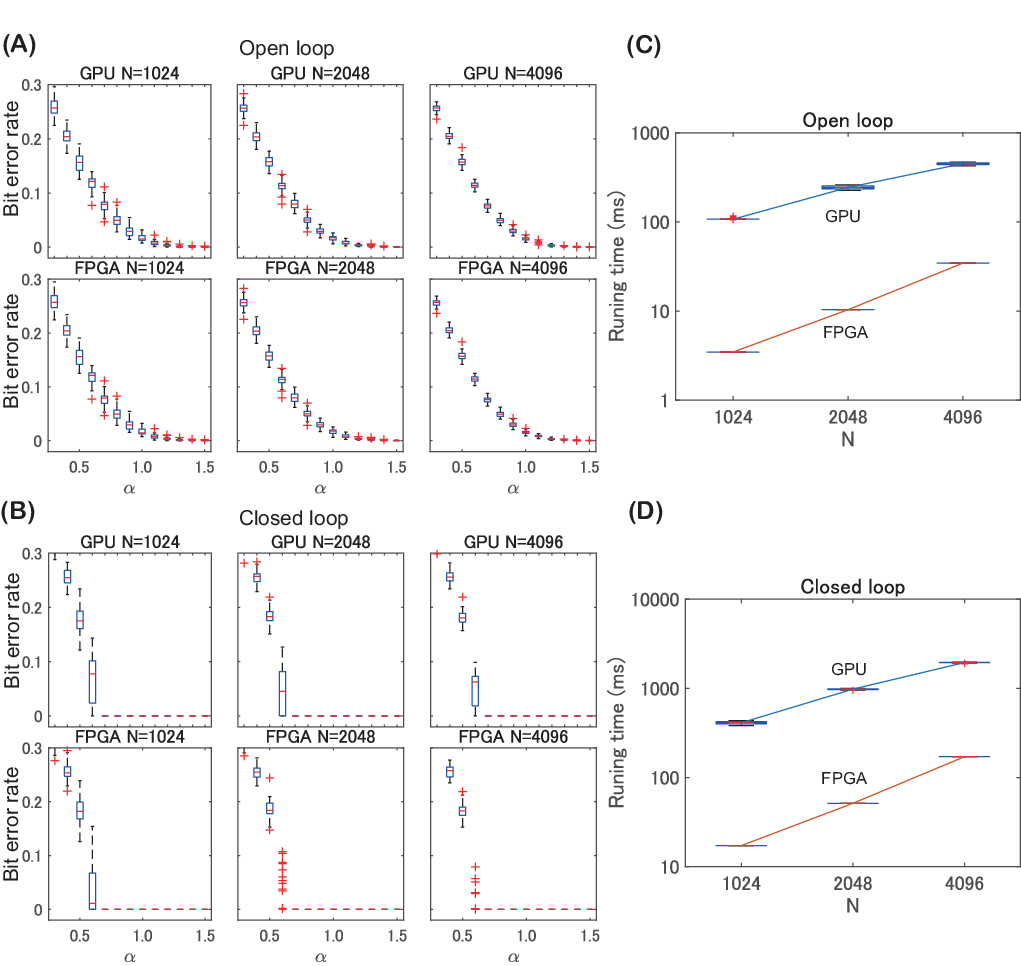}
\end{center}
\caption{Comparison of decoding performance and running time for the CDMA multi-user detectors on cyber CIMs implemented with GPU and FPGA. System sizes of these are $N=1024$, $N=2046$ and $N=4096$.
(A) Relationship between spreading rate $\alpha$ and bit-error rate for open-loop CIM on GPU and FPGA.
(B) Relationship between $\alpha$ and bit-error rate for closed-loop CIM on GPU and FPGA.
(C) Running time for open-loop CIM on GPU and FPGA.
(D) Running time for closed-loop CIM on GPU and FPGA.}
\label{fig.cdma_result}
\end{figure*}

\subsubsection{Comparison of decoding performance and running time}

We evaluated the decoding performance and running time of open-loop CIM and closed-loop CIM implemented on GPU and FPGA. We generated spreading code series from 50 different random seeds to obtain 50 samples of different coupling matrices $J$ and Zeeman terms. We then compared the decoding accuracy and running times for the open-loop CIM and closed-loop CIM on the parallel FPGA and GPU implementations, using the same 50 samples of coupling matrices $J$ and Zeeman terms. The parameter values used in this experiments are shown in Table \ref{table:Parameter}. Three system sizes were used: $N=1024$, $N=2046$ and $N=4096$.

Figure \ref{fig.cdma_result} (A) shows a box plot of the relationship between spreading rate $\alpha$ and bit-error rate for open-loop CIM on GPU and FPGA, and Figure \ref{fig.cdma_result} (C) shows the running times for each system size $N$. Figure \ref{fig.cdma_result} (B) and (D) similarly show the relationship between spreading rate $\alpha$ and bit-error rate, and running times for closed-loop CIM on GPU and FPGA.

As shown in Fig. \ref{fig.cdma_result} (A) and (B), with both FPGA and GPU, the bit-error rates were lower for closed-loop CIM than for open-loop CIM. This is due to the function of the CAC in closed-loop on FPGA and GPU, establishing the balance between the size of Zeeman and interaction terms, and further the destabilization of local minimum states improving the performance of finding an optimal solution \cite{RN1196,RN1153,RN1200,RN1158,RN1159,RN1143,RN1204}. As also can be seen in Fig. \ref{fig.cdma_result} (A), for open-loop CIMs on GPU and FPGA, the bit-error-rate distributions of both were the same for all values of $N$. On the other hand, as shown in Fig. \ref{fig.cdma_result} (B), for closed-loop CIM, even though the same samples of the spreading code series were used for FPGA and GPU, for all values of $N$ the bit-error rate for FPGA at $\alpha=0.6$ was lower than the value for GPU. As discussed in the previous section, this difference could be because our FPGA system does not conform to rounding rules of IEEE 754 FP32. Due to the increased dynamical complexity inherent in closed-loop CIM resulting from CAC, they exhibit sensitivity to rounding errors in numerical calculations. Consequently, differing solution trajectories were observed between FPGA and GPU, leading to disparities in decoding performance.

As shown in Fig. \ref{fig.cdma_result} (C) and (D), for both open-loop CIM and closed-loop CIM, running time on GPU was 11-times longer than on FPGA. Table \ref{table:cdma_runtime} summarize the median values of running times on FPGA and GPU and their ratios for each system size $N$. The running times of closed-loop CIM were around five-times longer than those of open-loop CIM for both FPGA and GPU because the number of repetitions $N_{step}$ was five-times higher. The running times on the FPGA for each system sizes $N$ coincide with the number of cycles estimated using Eq. \ref{eq.FPGA_CIM_cycle} (Table \ref{table:EST_CYC}) divided by the clock frequency (30 MHz). Comparing the time ratios in Table \ref{table:cdma_runtime}, the differences in running times for FPGA and GPU were more noticeable for smaller values of $N$, suggesting that there is more processing overhead for the GPU case.

\begin{table}[t]
    \caption{Running-time median value of GPU and FPGA processing CDMA multi-user detector}
 \label{table:cdma_runtime}
 \centering
  \begin{tabular}{llll}
   \hline
     & N=1024 & N=2048 & N=4096 \\
   \hline \hline
{\bf Open-loop CIM} &   &   &  \\
FPGA [$ms$] & 3.47 & 10.35 & 34.49 \\
GPU [$ms$] & 107.20 & 243.42  & 449.85 \\
Time ratio & 30.89 & 23.52 & 13.04\\
\hline
{\bf Closed-loop CIM} &   &   &  \\
FPGA [$ms$] & 17.20 & 51.35 & 171.09 \\
GPU [$ms$] & 411.50 & 979.81 & 1947.52 \\
Time ratio & 23.92 & 19.08 & 11.38 \\
   \hline
  \end{tabular}
\end{table}

\subsection{L0RBCS (QUBO Hamiltonian)}

In this section we focus on L0RBCS as an example of the QUBO Hamiltonian with Zeeman terms \cite{RN1141,RN1142,RN1204}. In the section {\bf Implemented Algorithms}, we described the L0RBCS optimization problems handled in this research. As shown in Fig. \ref{fig:algorithm_relation} and Algorithm \ref{alg6}, our architecture performs L0RBCS optimization by alternating between either open-loop CIM (Algorithm \ref{alg1}) or closed-loop CIM (Algorithm \ref{alg3}) and Jacobi SOR (Algorithm \ref{alg5}). When performing CS with open-loop CIM (Algorithm \ref{alg1}) and Jacobi SOR (Algorithm \ref{alg5}) it is called open-loop CS \cite{RN1141}, and when performing CS with closed-loop CIM (Algorithm \ref{alg3}) and Jacobi SOR (Algorithm \ref{alg5}) it is called closed-loop CS \cite{RN1142,RN1204}.

We compared the reconstruction accuracy and running times for open-loop CS and closed-loop CS on FPGA and GPU, using the same observation matrices and observed signals.

\begin{figure*}[htbp]
\begin{center}
\includegraphics[width=150mm]{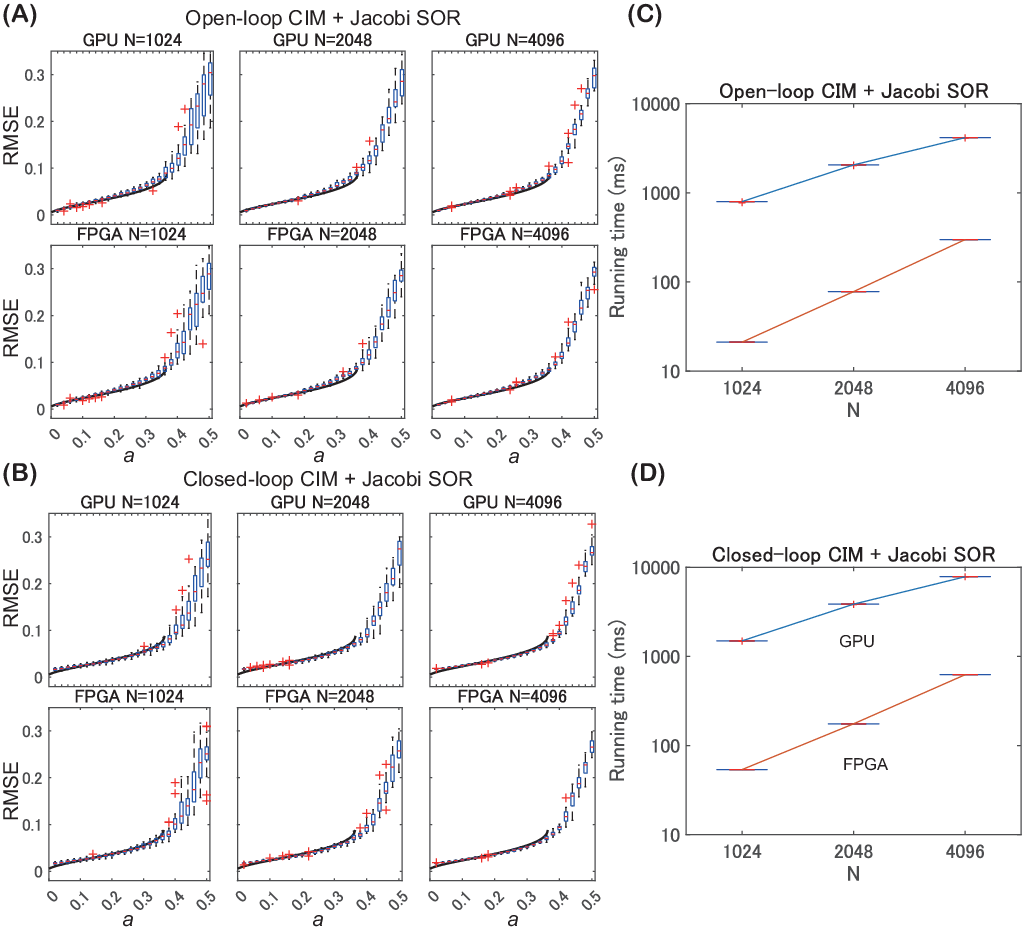}
\end{center}
\caption{Comparison of random signal reconstruction performances and runnning times for L0RBCS on cyber CIMs implemented with GPU and FPGA. For a comparison of cyber CIMs implemented with GPU and FPGA, the same randomly-generated observation matrices and observation signals were set in the both, and these performances and calculation times were evaluated. System sizes of these are $N=1024$, $N=2046$ and $N=4096$.
(A) Relationship between sparseness $a$ and RMSE for open-loop CS on GPU and FPGA.
(B) Relationship between $a$ and RMSE for closed-loop CS on GPU and FPGA.
The solid-black lines in (A) and (B) are RMSE at a ground state for each $a$ predicted with statistical mechanics analysis.
(C) Running time per single reputation of alternating optimization in open-loop CS on GPU and FPGA.
(D) Running time per single reputation of alternating optimization in closed-loop CS on GPU and FPGA. The compression rate for all figures was $\alpha=0.8$.}
\label{fig.stat_result}
\end{figure*}

\subsubsection{Evaluation of performance with randomly generated observation matrices and observed signals}

Under conditions where the ground state can be verified through statistical mechanics, we checked whether the reconstruction by open-loop and closed-loop CSs on GPU and FPGA matched the theoretically predicted ground state, and evaluated the running times on the different hardware-implemented CSs. 

According to conditions for realizing statistical mechanics analysis for L0RBCS in our previous research \cite{RN1141}, we configured variables in the observation model defined in Eq. \ref{eq.cs1} as follows. Each entry $A_i^\mu$ of the observation matrix $A$ was randomly generated from independent identical normal distribution, satisfying $\left<A_i^\mu\right>=0$ and $\left<A_i^\mu A_j^\nu\right>=\frac{1}{M}\delta_{ij}\delta_{\mu\mu}$. For the sparseness $a$, $aN$ elements of source signal $x$ were generated randomly from independent identical normal distribution with a mean of $0$ and a standard deviation of $1$, and other elements of $x$ were set to zero. For these experiments, the observation noise standard deviation $\zeta$ was set to $0.05$ and the compression rate $\alpha$ was set to $0.8$. Under these conditions, we used 16 different random seeds to randomly generate observation matrices, source signals and observation noise and then created observation signals according to Eq. \ref{eq.cs1}. This produced 16 samples of coupling matrices $J$ and Zeeman terms according to Eq. \ref{eq.CS_zeeman}. For this experiment, we set $\Gamma_k=0$. We then compared the reconstruction accuracy and running times for open-loop CS and closed-loop CS on FPGA and GPU using the same samples of coupling matrices $J$ and Zeeman terms. The parameter values for this experiment are given in Table \ref{table:Parameter}. Three system sizes were used: $N=1024$, $N=2046$ and $N=4096$.

Figure \ref{fig.stat_result} (A) is a box plot showing the relationship between sparseness $a$ and RMSE for open-loop CS on GPU and FPGA, and Fig. \ref{fig.stat_result} (C) is a box plot of running time for one repetition of alternating optimization for each system size $N$. Similarly, Figures \ref{fig.stat_result} (B) and (D) show the relationship between $a$ and RMSE and the running time per repetition for closed-loop CS on GPU and FPGA. The solid black lines in Fig. \ref{fig.stat_result} (A) and (B) are predicted RMSE values in the ground state for each $a$, obtained using statistical mechanics analysis.

As shown in Fig. \ref{fig.stat_result} (A) and (B), for both open-loop CS and closed-loop CS, the RMSE distributions for all $N$ are the same on both GPU and FPGA. In this experiment, unlike with the CDMA multi-user detector, we were not able to confirm a difference in reconstruction accuracy between FPGA and GPU for closed-loop CS. Also, as shown in Fig. \ref{fig.stat_result} (A) and (B), when $0.2<a<0.35$, the RMSE for closed-loop CS on both FPGA and GPU was closer to the theoretically-predicted ground-state RMSE than it was with open-loop CS \cite{RN1142,RN1204}. This could be due to the function of the CAC in closed-loop CIM on FPGA and GPU, which destabilizes local minimum states, leading to a solution closer to the ground state \cite{RN1196,RN1153,RN1200,RN1158,RN1159,RN1143,RN1204}.

Figure \ref{fig.stat_result} (C) and (D) show that for both open-loop CS and closed-loop CS, GPU running times were more than 12-times longer than FPGA times. Table \ref{table:L0CS_random_runtime} shows the median values of running time on FPGA and GPU and their ratios for each system sizes $N$. For both FPGA and GPU, running times for closed-loop CS were more than twice as long as for open-loop CS, because the total number of iterations of CIM and Jacobi SOR, $N_{step}$, is about double. The running times on the FPGA for each system sizes $N$ coincide with the number of cycles estimated using Eq. \ref{eq.FPGA_CIM_cycle} (Table \ref{table:EST_CYC}) divided by the clock frequency (30 MHz). Comparing the time ratios in Table \ref{table:L0CS_random_runtime}, the differences in running times for FPGA and GPU were more noticeable for smaller values of $N$, suggesting that there is more processing overhead for the GPU case.

\begin{table}[t]
 \caption{Running-time median value of GPU and FPGA processing L0RBCS with random observation matrix and random source signal}
 \label{table:L0CS_random_runtime}
 \centering
  \begin{tabular}{llll}
   \hline
     & N=1024 & N=2048 & N=4096 \\
   \hline \hline
{\bf Open-loop + Jacobi} &   &   &  \\
FPGA [$ms$] & 21.10 & 77.97 & 299.47 \\
GPU [$ms$] & 796.91 & 2056.85  & 4164.35 \\
Time ratio & 37.77 & 26.38 & 13.91 \\
\hline
{\bf Closed-loop + Jacobi} &   &   &  \\
FPGA [$ms$] &  53.72 & 175.34 & 623.89  \\
GPU [$ms$] & 1493.55 & 3855.67 & 7830.34 \\
Time ratio & 28.18 & 21.99 & 12.55\\
   \hline
  \end{tabular}
\end{table}

\begin{figure*}[htbp]
\begin{center}
  \includegraphics[width=150mm]{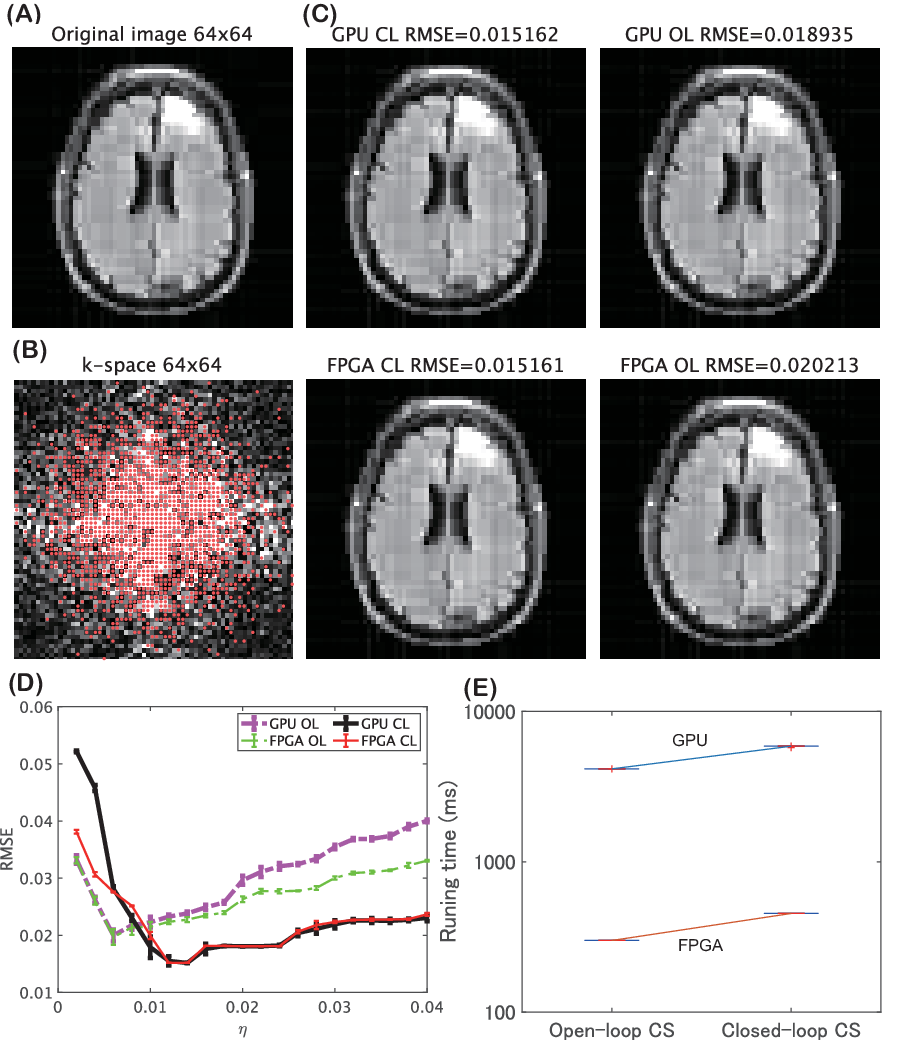}
\end{center}
\caption{Comparing MRI image reconstruction performances and running times for open-loop CS and closed-loop CS on GPU and FPGA. Using 64x64 pixel (N=4096) MRI data, we evaluate performance of open-loop CS and closed-loop CS on GPU and FPGA.
(A) 64x64 pixel original image. This image is composed of 18.2$\%$ of all Haar bases. Thus, sparseness $a=0.182$.
(B) k-space data for original image. The red points are random sampling points obeying a normal distribution. Undersampling was executed at a compression rate of $\alpha=0.4$.
(C) Example of images reconstructed with open-loop CS and closed-loop CS on FPGA and GPU from observed data obtained through random under-sampling in (B). Minimum error images and corresponding RMSE under each of the different hardware implemented the CSs. CL: closed-loop CS. OL: open-loop CS. Thresholds for minimum RMSE in each are $\eta=0.006$ for GPU OL,  $\eta=0.006$ for FPGA OL, $\eta=0.014$ for GPU CL and $\eta=0.014$ for FPGA CL.
(D) The relationship between RMSE and threshold $\eta$ in each case calculated with the same 16-samples of the random sampling point set. (E) Running time per repetition of alternating optimization in each case.}
\label{fig.MRI_result}
\end{figure*}

\subsubsection{Performance evaluation with MRI data}

Using MRI data, we evaluate the performance of open-loop CS \cite{RN1141} and closed-loop CS \cite{RN1204} on GPU and FPGA. The image used for this evaluation is a cross-sectional image from the fastMRI data set \cite{RN1203}, converted to a 64x64 pixel size. We applied a Haar-wavelet transform (HWT) to the data, and selected the top 18.2$\%$ components with large HWT coefficients to create a signal spanned by Haar basis functions with a sparseness of 0.182 (Fig. \ref{fig.MRI_result} (A)). We applied a discrete Fourier transform (DFT) to the image to obtain the k-space data shown in Fig. \ref{fig.MRI_result} (B). From this k-space data, we sampled 40$\%$ of the k-space data at random red points obeying a two-dimensional normal distribution. Thus, we performed random undersampling with a compression rate of $\alpha=0.4$. In this experiment, we generated 16 sets of random undersampling points from 16 different random seeds. The red scattered points in Fig. \ref{fig.MRI_result} (B) show an example of the random sampling points. 

Based on the above conditions, we performed L0RBCS optimization on the Haar space. The coupling matrix $J$ and Zeeman terms for compressed sensing as defined in Eq. \ref{eq.CS_zeeman} were set as follows.
\begin{eqnarray}
   A &=& SF\Psi^T, \ \ \Gamma_1 = \Delta_h\Psi^T, \ \ \Gamma_2 = \Delta_v\Psi^T, \nonumber \\
   J &=& -A^T A + \gamma \Gamma_1^T\Gamma_1 + \gamma \Gamma_2^T\Gamma_2, \ \ g = A^T y. \nonumber 
\end{eqnarray}
Here, $\Psi$ is the HWT matrix and $F$ is the DFT matrix. $S$ is the matrix performing random undersampling as described above, which is different for each random seed. $\Delta_h$ and $\Delta_v$ are respectively the second-derivative matrices for the vertical and horizontal directions. $\gamma$ is a L2-norm regularization parameter, set to $\gamma=0.0001$ for this experiment. $y$ is the observed signal, created according to the observation model in Eq. \ref{eq.cs1}. For this experiment, we set the observation noise to zero ($\zeta=0$). The initial state for open-loop CS and closed-loop CS on FPGA and GPU was given by LASSO \cite{RN1212}. We also set threshold values to a fixed value, $\eta_{int} = \eta_{end} = \eta$, during the repetitions of alternating optimization. These experimental conditions are the same as we used in our previous research \cite{RN1141,RN1142,RN1204}.

Figure \ref{fig.MRI_result} (C) shows examples of images reconstructed from observed data obtained through random undersampling as shown in Fig. \ref{fig.MRI_result} (B). It shows the minimum-error images reconstructed by closed-loop CS and open-loop CS on GPU and FPGA, along with their RMSE. The threshold values $\eta$ minimizing the RMSE in each case are given in the caption for Fig. \ref{fig.MRI_result}. Figure \ref{fig.MRI_result} (D) shows the relationship between the RMSE of the reconstructed image under each of the conditions and the threshold value $\eta$ used, for each of the 16 sets of random sampling points. The error bars show the mean and standard deviation of the RMSEs for the 16 sets.

For open-loop CS on both GPU and FPGA, RMSE was smallest near $\eta=0.006$, and for closed-loop CS, RMSE was smallest near $\eta=0.014$. The minimum RMSE value was lower for closed-loop CS than for open-loop CS for both FPGA and GPU implementations. This could be due to the function of the CAC in closed-loop CIM on FPGA and GPU, which destabilizes local minimum states, leading to a solution closer to the ground state. For open-loop CS, there was a difference in RMSE between GPU and FPGA implementations when $\eta>0.01$, while for closed-loop CS there was a difference between GPU and FPGA when $\eta<0.012$. These differences could be because our FPGA system does not conform to rounding rules of IEEE 754 FP32. In contrast to the CDMA multi-user detector results, there was also a difference between. GPU and FPGA results with open-loop CS. Figure \ref{fig.MRI_result} (D) shows that the difference between GPU and FPGA with open-loop CS occurred when RMSE was around 0.01 and above. It has been reported that when the RMSE is high, the system is in a high frustration state, and in such states the complexity of system behavior increases \cite{RN1210}. Thus, the difference in rounding rules between GPU and FPGA could have resulted in the difference in calculation results for open-loop CS as well.

Figure \ref{fig.MRI_result} (E) shows a box plot of the running time per repetition for  alternating optimization under the various conditions. For both open-loop CS and closed-loop CS, running times were at least 12-times longer for the GPU than for the FPGA implementation. Table \ref{table:L0CS_MRI_runtime} summarizes the median values of running times on FPGA and GPU and their ratios. The FPGA running times correspond with the number-of-cycle values estimated using Eq. \ref{eq.FPGA_CIM_cycle} (Table \ref{table:EST_CYC}) divided by the clock frequency (30 MHz). 

\begin{table}[t]
 \caption{Running-time median value of GPU and FPGA processing L0RBCS for MRI data with 64x64 pixels (N=4096)}
 \label{table:L0CS_MRI_runtime}
 \centering
  \begin{tabular}{lll}
   \hline
     & Open-loop + Jacobi & Close-loop + Jacobi \\
   \hline \hline
FPGA [$ms$] & 299.47 & 453.14 \\
GPU [$ms$] & 4152.62 & 5876.84 \\
Time ratio & 13.87  & 12.97 \\
   \hline
  \end{tabular}
\end{table}

\section{Discussion}

\subsection{Architecture versatility and its necessity}

For the architecture we have developed, three different algorithms: open-loop CIM, closed-loop CIM and Jacobi SOR can be executed in the same modules. As described in the section {\bf FPGA architecture}, the calculations of these algorithms are mainly performed by CAL$\_$CSR for TE calculation, CAL$\_$H for the local field calculation, and CTL$\_$TO that controls these modules. CAL$\_$CSR consists of two adders, two multipliers, a square-root module, a random number generator and FFs to temporarily store the calculated results. CAL$\_$H has 2048 MAC PEs. CTL$\_$TO switches the operations of these arithmetic modules in CAL$\_$CSR and CAL$\_$H every cycle according to the sequence control code stored it to proceed with the calculations. Thus, by switching the sequence control code, three different types of algorithms can be executed in the same modules. CAL$\_$CSR has the arithmetic modules needed for the calculation of the SB, so our FPGA system could also perform the SB calculation by rewriting the sequence control code. Thus, the architecture we have developed is more versatile than architectures developed in preceding researches \cite{RN1050,RN1139,RN1135,RN1192,RN1133}.

As research progresses, the CIM models and other systems are subject to frequent updates. However, it is difficult to redesign FPGA architecture to adapt to each update. There is a need for an architecture designed to maintain versatility to adapt to the update and also provide computing speeds that are superior to using GPUs. Our FPGA system can support both open-loop CIM, which was proposed early in CIM research, and closed-loop CIM recently used, and furthermore is superior to GPUs in calculation speed by a factor of ten or more for both models.

\subsection{Utility of discrete and continuous-valued interaction}

For almost all Ising algorithms, the most time-consuming part of the computation is calculating the local field, which requires the multiplication of $N\times N$ coupling matrices with $N$-dimensional spin-state vectors. By using discrete-valued interactions in FPGA implementation, it is possible to reduce the memory and logic elements resulting in faster clock speed and larger degree of parallelism. And there have been prior implementations of large-scale FPGA-based systems with discrete-valued interactions \cite{RN1050,RN1139,RN1135,RN1192,RN1133}.

We have focused on the Hopfield model as an example of an Ising model, and evaluated the effects of using discrete-valued interactions \cite{RN1147,RN1148}. We used statistical mechanics to evaluate the degradation in performance of a conventional Ising model and a Wigner-SDE CIM model when the continuous-valued interaction $J_{ij}$ determined by Hebb rule is discretized to three values, $\{-1, 0, 1\}$. The results of the analysis showed that recalled patterns degraded slightly, and the critical memory capacity declined by 30$\%$. When using four-bit discrete-valued interactions, there was almost no performance degradation. The reason is that the effect of discretization of the interaction is equivalent to the quenched Gaussian noise due to the central limit theorem. In memory retrieval phase, this noise component is relatively small in the local field compared to the ferromagnetic component, so this noise component does not strongly prevent pattern retrieval, and its influence on pattern retrieval is small. This finding of the Hopfield model suggests that models with strong ferromagnetic properties could be less affected by the interaction discretization.

In contrast with the Hopfield model, for signal processing models such as compressed sensing \cite{RN1141,RN1142}, CDMA multi-user detector \cite{RN1144} and MIMO \cite{RN1149,RN1152,RN1151}, the interactions $J_{ij}$ of these models are anti-Hebb type with opposite sign. (See Eqs. \ref{eq. anti-hebb-cdma} and \ref{eq.CS_zeeman}). For these signal processing models, the Zeeman term can work as the matched filter, and the anti-Hebb-type mutual interaction term plays a role in removing cross-talk noise in the matched filter realized in the Zeeman term. If the interaction for these models is discretized, it would not be possible to eliminate cross-talk noise, which would greatly degrade performance. Preliminary experiments have shown that for compressed sensing, discretizing interactions using a four-bit representation results in a large decline in signal recovery performance. As such, it is essential to use continuous-valued interactions for these signal processing models.

There are some special combinatorial optimization problems that can be represented as an Ising model with discrete-valued interactions. There are models that are not significantly effected by discretization of interactions, like the Hopfield model described above, but these are special cases. Considering system versatility for application to real problems,  continuous-valued interactions are preferable.

\subsection{Problems with J$\_$MUX wiring delay and operation clock}

The FPGA we are using is the Xilinx ALVEO U250, and its maximum clock frequency is 300 MHz. However, we are operating it at 30 MHz rather than this maximum frequency. Our operating clock is also much lower than the 270 MHz used for the SB FPGA implementation (Table \ref{table:FPGA_CIM}). The reason for reducing the clock frequency in this way was due to wiring delay in J$\_$MUX.

It was not possible to store $4096\times 4096$ matrix $J$ with FP32 presentation in internal memory, so the upper triangular part of the matrix $J$ was stored as shown in Fig. \ref{fig.FPGA_arc} (D). J$\_$MUX is a multiplexer that fills-in the symmetric matrix entries from the data in memory. In this manner, we can implement the full-coupling local field calculation under full-coupling with FP32 representation up to the system size of $N=4096$ on a single FPGA. However, wiring delay in J$\_$MUX resulted in a bottleneck to operating speed. 

To increase the operating clock, it is necessary to reduce the delay by improving the efficiency of J$\_$MUX. Alternatively, if an FPGA unit with larger internal memory than Xilinx ALVEO U250 becomes available and can store $4096\times 4096$ matrix $J$ in the internal memory, J$\_$MUX will no longer be needed, making clock frequency over 100 MHz possible. 

\subsection{Improving calculation speed by increasing parallelism}

As described in Section 4.4, the number of parallels in our single FPGA system is $P_b\times P_r=64$ and the number of parallels for the single FPGA SB implementation was  $P_b\times P_r=256$ \cite{RN1139,RN1135}. Accordingly, for our FPGA implementation, the number of cycles per step for calculations of the local field and the TE was four times higher than for the single FPGA SB. The reason for lower parallelism was that we implemented CAL$\_$H using FP32 MACs, which require more logic elements than those needed for binary MACs. It would be possible to increase the degree of parallelism by ASIC implementation or if FPGA units with greater logic capacity became available. An FPGA cluster consisting of multiple nodes could also be used to increase the degree of parallelism. As shown in Fig. \ref{fig.matrix_para} (A), the matrix calculation can be divided into blocks and the calculations for each block can be shared among multiple FPGAs. TE operations can also be parallelly calculated using partial local fields calculated on each FPGA. Whenever one step of the calculation is performed, the spin state updated at each node is shared between all nodes through inter-FPGA communication. A previous research on multi-node FPGA implementation for SB has shown that this inter-FPGA communication can be achieved in the same number of cycles as the number of nodes, and thus the communication time does not become a serious bottleneck in this parallel computation \cite{RN1135}. Highly efficient inter-FPGA communication methods, which is done during operations such as MAC, have also been proposed \cite{RN1192}. Our system could also be clustered using similar methods.

\section{Conclusions}

In this research, we have developed a highly versatile cyber CIM implemented in an FPGA.
Using the architecture we have developed, we can execute the open-loop CIM that was proposed in early CIM research, the closed-loop CIM that has been proposed recently, as well as Jacobi SOR on a single module. By simply rewriting the sequence control code for the calculation control module, various algorithms can be executed, including the SB. In contrast with large-scale FPGA implementations in earlier research, in which interactions are represented with binary or ternary values, our system uses interactions with FP32 representation. Furthermore, our system uses a FP32 representation for Zeeman terms, as opposed to earlier studies, which used a binary representation. System sizes implemented on a single FPGA are  $N=1024$, $N=2048$ and $N=4096$. This is comparable to the single-FPGA SB implementation with the binary-interaction, which had $N=4096$. Our system has $1/4$ degree of parallelism of the single-FPGA SB implementation, so it requires approximately four-times the number of cycles for one step including the local field and TE calculations. Despite this, it is a highly versatile system, and it is capable of solving problems that cannot be solved with earlier FPGA systems, such as CDMA multi-user detectors and L0RBCS, at greater speed than using a GPU (approximately more than 10 times faster). Calculation speed could also be further improved by increasing the degree of parallelism through clustering and so on.

\section*{Acknowledgements}

This research was conducted with support from NTT Research Inc. and an NSF CIM Expedition award (CCF-1918549). We would also like to thank Mr. Naohiro Iida from Ryoyo Electro Corporation for his efforts with tenders, contracts, adjusting and reaching agreement among the groups.

\bibliographystyle{IEEEtran}
\bibliography{reference1}

\end{document}